\documentclass[showpacs,preprintnumbers,amsmath,amssymb,nofootinbib]{revtex4}

\def\beq{\begin{equation}}
\def\eeq{\end{equation}}
\def\beqa{\begin{eqnarray}}
\def\eeqa{\end{eqnarray}}

\def\za{\alpha}
\def\zb{\beta}
\def\lsim{\mathrel{\raise.3ex\hbox{$<$\kern-.75em\lower1ex\hbox{$\sim$}}} }
\def\gsim{\mathrel{\raise.3ex\hbox{$>$\kern-.75em\lower1ex\hbox{$\sim$}}} }
\usepackage{color}
\usepackage{latexsym}
\usepackage{graphicx}
\usepackage{amssymb}
\usepackage{amsmath}

\begin{document}
\preprint{{\vbox{\hbox{NCU-HEP-k029} \hbox{Dec. 2008} 
\hbox{ed. Jun 2009}}}}

\vspace*{.5in}
\title{Leptonic Radiative Decay in
  Supersymmetry without R parity}

\vspace*{.5in}

\author{Chien-Yi Chen}
\email{chienyic@andrew.cmu.edu} \affiliation{Department of
Physics, Carnegie Mellon University, Pittsburgh, Pennsylvania
15213, USA \\ and Department of Physics, National Central University,
Chung-Li, Taiwan 32054}
\author{Otto C. W. Kong}
\email{otto@phy.ncu.edu.tw} \affiliation{Department of Physics and
Center for Mathematics and Theoretical Physics, National Central
University, Chung-Li, Taiwan 32054}

\vspace*{1in}

\begin{abstract} We present a detailed analysis together with exact
numerical calculations on one-loop contributions to the branching
ratio of the radiative decay of $\mu$ and $\tau$ , namely $\mu \to
e\, \gamma$, $\tau \to e\,\gamma$, and $\tau \to\mu \gamma$ from
supersymmetry without R parity, focusing on contributions involving
bilinear couplings. A numerical study is performed to obtain
explicit bounds on the parameters under the present experimental
limit. We present, and use in the calculation, formulas for exact
mass eigenstate effective couplings. In this sense, we present an
exact analysis free from approximation for the first time.
After comparing our results against the closest early analysis, we
discovered a major difference in resulted constraints on
some ${\mu_i^*}\,{B_j}$ combinations. Constraints from neutrino
masses on the parameters were considered. Our result indicates that
the branching ratio measurement on $\mu \to e\, \gamma$ down to
$10^{-13}-10^{-14}$ and beyond, as targeted by the MEG experiment,
has a chance of observing decay from the R-parity violating
scenario.
%%%%%%%%%%%%%%%%%%%%%%%%%%%%
\end{abstract}


\maketitle

\newpage

%\tableofcontents
%%%%%% Section 1 %%%%%%%%%%%%%%%%%%%%%%%%%%%%%%%%%%%%%%%%%%%%%%%%%%%%%
\section{Introduction}
Recent neutrino experiments have demonstrated that neutrinos
change flavor as they travel from source to detector, a
phenomenon consistent with the hypothesis of neutrino oscillation.
All that contributes evidence for neutrino masses and lepton-flavor 
violation (LFV) and provides the first definite
experimental clue for physics beyond the standard model (SM). Many
extensions of the SM predict a certain amount of LFV in relation
to neutrino mass generation or otherwise. Important criterion for a
viable model is giving acceptable neutrino mass spectrum while staying
within experimental limits of LFV. Apart from the soft terms within
the minimal supersymmetric standard model (MSSM),
both LFV and neutrino masses are indeed
forbidden by {\it ad hoc} discrete symmetry --- the R parity.
Note that soft terms by themselves still conserve total lepton
number, and hence do not generate neutrino masses. In the
supersymmetric standard model without R parity imposed, there is
however an important source of LFV and neutrino masses. A major part
of this comes simply from the R-parity violating (RPV) terms in
the superpotential, though RPV soft (supersymmetry breaking) terms
are also of interest. The latter too often escapes notice.

The best evidence of supersymmetry (SUSY) would obviously be the
discovery of SUSY particles in the collider machines. However,
processes such as the leptonic radiative decays can serve as
alternative ways to test SUSY, complementary to the direct SUSY
particle searches. Although these processes have not yet been seen so
far in present experiments, there are very stringent upper
bounds on their possible rates implying important constraints
on the new physics contributions. The present experimental upper
bounds of branching ratio for $\tau \to \mu \,\gamma$
\cite{Aubert:2005ye} , $\tau \to e\, \gamma$ \cite{Aubert:2005wa},
and $\mu \to e\, \gamma$ \cite{Brooks:1999pu} are
\[
Br(\tau \to \mu \gamma) < 6.8\times 10^{-8} \;,
\]
\[
Br(\tau \to e\,\gamma) < 1.1\times 10^{-7} \;,
\]
\[
Br(\mu \to e\, \gamma) < 1.2\times 10^{-11} \;.
\]
The muon radiative decay reaction $\mu \to e\, \gamma$ has been
the focus of most attention due to the experimental bound being
much stronger.
This bound will probably be improved in the future. The MEG
experiment Ref.\cite{Maki:2008zz}, which searches for $\mu \to e\,
\gamma$ decays down to $10^{-13}-10^{-14}$ branching ratio is now
in its final stage of preparation. The $\tau$ decays may also
be better probed in future facilities.

The recent studies on radiative decays from other models such
as little Higgs with T parity
\cite{Choudhury:2006sq,Blanke:2007db} and the
SUSY grand unified theories
(GUT)\cite{Calibbi:2006nq,Calibbi:2006ne} also give some
interesting results on the lepton-flavor violation processes. For example, in
the little Higgs models with T parity, the presence of
new flavor violating interactions and mirror leptons containing
masses of order 1TeV can enhance these processes to the level of the
present experimental limit. In the SUSY GUT model, Ref.\cite{Calibbi:2006nq}
discusses the complementarity between lepton-flavor
violation and Large Hadron Collider (LHC) experiments in
probing the SUSY GUT. They found that the LFV experiments
have strong capabilities to detect SUSY induced LFV, in some cases
even outreaching the LHC. In Ref.\cite{Calibbi:2006ne}, the
authors study the correlation between ${U_{e3}}$ and the $Br(\mu
\to e\, \gamma)$ in the context of a SUSY SO(10) framework. They
find that taking running effects into account leads to a constant
enhancement of the value of ${U_{e3}}$ at the high scale, bringing
$\mu \to e\, \gamma$ into the realm of MEG for SUSY parameter
space regions which were previously excluded without inclusion
of such running.

The study we present in this paper analyzes branching ratios
that can be generated for all processes in the context of
the generic supersymmetric standard model (GSSM), {\it i.e.} SUSY
without R parity \cite{Kong:2002hb}. If one simply takes the
minimal supersymmetric field spectrum of the SM and imposes
nothing more than gauge symmetries while admitting soft SUSY
breaking, the generic supersymmetric standard model would be
obtained. Thus, the GSSM is the complete theory of SUSY without R
parity, where all kinds of RPV terms are admitted without bias.
Assuming SUSY, it is at least conceptually, the simplest model to
accommodate neutrino mixing and oscillations. We work within the
framework of single-VEV parametrization
(SVP)\cite{Kong:2002hb,Bisset:1998bt,Bisset:1998hy}, which is an
optimal choice of flavor basis that helps guarantee a
consistent and unambiguous treatment of all kinds of admissible
RPV couplings and to maintain a simple structure for RPV effects
on tree-level mass matrices for all states including scalars and
fermions.

The experimental bound on the branching ratio of these leptonic
radiative decays is used to constrain the model parameter space,
particularly the RPV part. Under constraints by the present
experimental upper bound of branching ratios for $\mu \to e\,
\gamma$ , $\tau \to e\, \gamma$ and $\tau \to \mu \,\gamma$, we
obtain the allowed region of the RPV parameter spaces. We give
complete one-loop formulas for the type of contributions to the
branching ratio of three leptonic radiative decays. We present
numerical analysis of these contributions from all possible
combinations of RPV parameters. Besides the more familiar
${\mu_k^*}\,{\lambda_{kij}}$ \footnote{The interesting kind of
combination of bilinear and trilinear RPV parameters
contributing to flavor violations through scalar mass mixings
\cite{as5} or a one-loop diagram \cite{Cheung:2001sb,korean}, and
analogous one-loop dipole moment \cite{edms} were published a few
years ago. More recently, similar contributions to radiative B
decays were also published \cite{rishi}.}, there are a list of
combinations of type ${B_k^*}\,{\lambda_{kij}}$,
${\mu_i^*}\,{\mu_j}$, and ${B_i^*}\,{\mu_j}$. A similar analysis
on the $\mu \to e\, \gamma$ process has been reported in
Ref.\cite{Cheung:2001sb} in 2001 
\footnote{There have been numerous
studies on similar processes from various versions or
limited models of R-parity violation in the literature. Most of
the model assumptions look {\it ad hoc}. We have no intention of
reviewing all of that here. However, an early study on $\mu \to e\,
\gamma$ from softly broken R parity \cite{Lee:1984kr} should
particularly be mentioned. Another particularly noteworthy paper
on the topic is given by Ref.\cite{deCarlos:1996aa}.}. 
The present
work differs from Ref.\cite{Cheung:2001sb} in a few important
ways. The present work is based on using new formulas of exact
mass eigenstate couplings to calculate the one-loop diagrams,
while in Ref.\cite{Cheung:2001sb} the authors
only used electroweak states (${l}_i^{\mp}$'s) as an
approximation for physical particles of external legs to
the loop (the charge leptons). The latter amounts to neglecting
the Higgsino and wino components of the decaying and product
charged leptons. Therefore, the current analysis is an improvement
or completion of the work reported in Ref.\cite{Cheung:2001sb}. In
particular, we find that the constraints one can obtain on some of
the ${\mu_i^*}\,{B_j}$ type parameter combinations have very
substantial improvement. This is indeed the first exact
calculation of processes within the model at the one-loop
level presented, free from any approximation of the type. In
addition, our results on the other two, $\tau$ decays, processes
have not been available in previous literature. We also increase
the value of the $\mu_{\scriptscriptstyle 0}$ parameter
(corresponding to the MSSM $\mu$ parameter) used from 100 to
135 GeV to accommodate the updated lightest chargino
$(\tilde{\chi}^{\pm}_{\scriptscriptstyle 1})$ lower mass limit of
approximately 104 GeV \cite{Eidelman:2004wy}. In the sense
explained above, the paper is somewhat of a sequel to
Ref.\cite{Cheung:2001sb}, where we draw comparison when
relevant. However, it can also be read just on its own. Readers
who want to do so may simply neglect statements matching analysis
and results here with that of the latter. One catch though is that
we focus our discussion and result presentations on
interesting results we get beyond that of
Ref.\cite{Cheung:2001sb}, only briefly summarizing features and
results that are essentially well explored in the latter
reference.

As experimental evidence for neutrino masses has become quite well
established, we also include in our analysis a brief comparison of
results from radiative decays with neutrino mass
bounds. Recombinations of RPV couplings typically contribute both to
the decays and neutrino masses, but with different dependence on
the other model parameter. Some RPV parameters, like the bilinear
ones, can give rise to a neutrino mass term alone. However, we do
not have solid evidence on the scale of the neutrino masses, only
$\Delta m^2$. And there are so many relevant RPV combinations for
both neutrino masses and radiative decay processes that it makes
a comprehensive and systematic analysis unrealistic
unless further assumptions are taken on the model
structure. Some parameter combinations may have a more important
role to play in a certain process while others may give dominant
contributions to neutrino masses. We are interested in
investigating and presenting generic results on model
parameters. Hence, we adopt a naive strategy on the interpretation
of neutrino mass bounds as naive upper bounds on the involved
parameter(s) in order not to give a neutrino mass term
contribution beyond the sub-eV scale. This may be a bit on the conservative side but is considered a reasonable strategy to be adopted. Note that this rough neutrino mass scale is not expected
to be pushed down, since the scale of $\Delta m^2$ is known from the oscillation experiments. On the other hand, we have only upper
bounds for the radiative decays which may be, and we believe
should be, probed with a lot better precision in the future. Our
radiative decay results will hence be useful references for the
future, even if they are no better than the naive neutrino mass
bounds. In the case of $\mu \to e\,\gamma$, for example, if the MEG
experiment pushes the bound on the branching ratio down to
$10^{-13}-10^{-14}$, our numerical results show that it can give a
stronger constraint on the ${\mu_1^*}\,{B_2}$ parameter
combination. To put it more interestingly, the neutrino mass
bounds do not rule out the possibility of seeing a $\mu \to
e\,\gamma$ at MEG coming from ${\mu_1^*}\,{B_2}$ of the RPV
supersymmetric model. In the case of ${\mu_{\scriptscriptstyle
2}^*}\lambda_{\scriptscriptstyle 212}$, the current bound is
actually already competitive.

This paper is organized as follows. In the next section, we
briefly summarize the main features of GSSM and also set our
notation. In Sec. III, we give the general exact formulas in the
basis of mass eigenstates without any approximation. The focus is
on the $\ell^{\!\!\mbox{ -}}_j  \to \ell^{\!\!\mbox{ -}}_i \,
\gamma$ amplitude from one-loop diagrams without colored
intermediate states. The two sections are included here to make the
paper self-contained, and set the notation to be used for the discussions
that follow. Note that Sec. III does include important results,
expressions for effective coupling among mass eigenstates involved, that
have not been published before. Our numerical results will be presented in
Sec. IV. We also compare the results obtained by using the exact
mass eigenstate couplings versus the previous calculations and
discuss the sources of the difference between them.  In addition,
we illustrate the effects of varying the input parameters on the
bounds. Finally, Sec. V will be devoted to the conclusions.

%%%%%%% Section 2 %%%%%%%%%%%%%%%%%%%%%%%%%%%%%%%%%%%%%%%%%%%%%%%%%%%%
\section{The Generic Supersymmetric Standard Model}
%%%%%%%%%%%%%%%%%%%%%%%%%%%%%%%%%%%%%%%%%%%%%%%%%%%%%%%%%%%%%%%%%%%%%%
We briefly describe the model here. Details of the formulation
adopted are elaborated on in Ref.\cite{Kong:2002hb}. The most general
renormalizable superpotential with the spectrum of minimal
superfields can be written as
\begin{equation}
W \!\! = \!\varepsilon_{ab}\left[ \mu_{\alpha}  \hat{H}_u^a
\hat{L}_{\alpha}^b + h_{ik}^u \hat{Q}_i^a   \hat{H}_{u}^b
\hat{U}_k^{\scriptscriptstyle C} + \lambda_{\alpha jk}^{\!\prime}
\hat{L}_{\alpha}^a \hat{Q}_j^b \hat{D}_k^{\scriptscriptstyle C} +
\frac{1}{2}\, \lambda_{\alpha \beta k}  \hat{L}_{\alpha}^a
 \hat{L}_{\beta}^b \hat{E}_k^{\scriptscriptstyle C} \right] +
\frac{1}{2}\, \lambda_{ijk}^{\!\prime\prime}
\hat{U}_i^{\scriptscriptstyle C} \hat{D}_j^{\scriptscriptstyle C}
\hat{D}_k^{\scriptscriptstyle C} \;  ,
\end{equation}
where  $(a,b)$ are $SU(2)$ indices, and $(i,j,k)$ are the usual family
(flavor) indices (going from $1$ to $3$). The $(\za, \zb)$ indices
are extended flavor indices going from $0$ to $3$. Note that
$\lambda$ is antisymmetric in the first two indices, as required
by the $SU(2)$ product rules, shown explicitly here with
$\varepsilon_{\scriptscriptstyle 12}
=-\varepsilon_{\scriptscriptstyle 21}=1$. Similarly,
$\lambda^{\!\prime\prime}$ is antisymmetric in the last two
indices from $SU(3)_{\scriptscriptstyle C}$, though color contents
are not shown here. Besides the superpotential, the Lagrangian
contains the gauge interaction part, including kinetic terms of
the matter superfields and a soft SUSY breaking part.

We take a definite flavor basis to write the model Lagrangian.
Such choice of parametrization is not unique. In the current case
of the GSSM, the scalar parts of the colorless electroweak doublet
superfields could bear vacuum expectation values (VEVs). We use a
parametrization called the SVP
advocated by our group since Ref.\cite{Bisset:1998bt}. A flavor
basis with only one among the $\hat{L}_\za$'s, designated as
$\hat{L}_0$, bearing a nonzero VEV is adopted. That is to say, the
direction of the VEV, or the Higgs field $H_d$, is singled out in
the four-dimensional vector space spanned by the $\hat{L}_\za$'s.
Explicitly, under the SVP, flavor bases are chosen such that (1)
$\langle \hat{L}_i \rangle \equiv 0$, which implies $\hat{L}_0
\equiv \hat{H}_d$; (2)  $y^{e}_{jk} (\equiv \lambda_{0jk}
=-\lambda_{j0k}) =\frac{\sqrt{2}}{v_{\scriptscriptstyle 0}}\,{\rm
diag} \{m_{\scriptscriptstyle 1}, m_{\scriptscriptstyle
2},m_{\scriptscriptstyle 3}\}$; (3) $y^{d}_{jk} (\equiv
\lambda^{\!\prime}_{0jk}) = \frac{\sqrt{2}}{v_{\scriptscriptstyle
0}}\,{\rm diag}\{m_d,m_s,m_b\}$; and (4)
$y^{u}_{ik}=\frac{\sqrt{2}}{v_{\scriptscriptstyle u}}\,
V_{\!\mbox{\tiny CKM}}^{\!\scriptscriptstyle T}\; {\rm
diag}\{m_u,m_c,m_t\}$, where $v_{\scriptscriptstyle 0}\equiv
\sqrt{2} \, \langle \hat{L}_0 \rangle$ and $v_{\scriptscriptstyle
u}\equiv \sqrt{2} \, \langle \hat{H}_{u} \rangle$.

The soft SUSY breaking part of the Lagrangian can be written as
follows \cite{as5,Kong:2002hb} : \beqa V_{\rm soft} &=&
\epsilon_{\!\scriptscriptstyle ab}
  B_{\za} \,  H_{u}^a \tilde{L}_\za^b +
\epsilon_{\!\scriptscriptstyle ab} \left[ \,
A^{\!\scriptscriptstyle U}_{ij} \, \tilde{Q}^a_i H_{u}^b
\tilde{U}^{\scriptscriptstyle C}_j + A^{\!\scriptscriptstyle
D}_{ij} H_{d}^a \tilde{Q}^b_i \tilde{D}^{\scriptscriptstyle C}_j +
A^{\!\scriptscriptstyle E}_{ij} H_{d}^a \tilde{L}^b_i
\tilde{E}^{\scriptscriptstyle C}_j   \,
\right] + {\rm h.c.}\nonumber \\
&&+ \epsilon_{\!\scriptscriptstyle ab} \left[ \,
A^{\!\scriptscriptstyle \lambda^\prime}_{ijk} \tilde{L}_i^a
\tilde{Q}^b_j \tilde{D}^{\scriptscriptstyle C}_k + \frac{1}{2}\,
A^{\!\scriptscriptstyle \lambda}_{ijk} \tilde{L}_i^a \tilde{L}^b_j
\tilde{E}^{\scriptscriptstyle C}_k \right] + \frac{1}{2}\,
A^{\!\scriptscriptstyle \lambda^{\prime\prime}}_{ijk}
\tilde{U}^{\scriptscriptstyle C}_i  \tilde{D}^{\scriptscriptstyle
C}_j \tilde{D}^{\scriptscriptstyle C}_k  + {\rm h.c.}
\nonumber \\
&&+
 \tilde{Q}^\dagger \tilde{m}_{\!\scriptscriptstyle {Q}}^2 \,\tilde{Q}
+\tilde{U}^{\dagger} \tilde{m}_{\!\scriptscriptstyle {U}}^2 \,
\tilde{U} +\tilde{D}^{\dagger} \tilde{m}_{\!\scriptscriptstyle
{D}}^2 \, \tilde{D} + \tilde{L}^\dagger
\tilde{m}_{\!\scriptscriptstyle {L}}^2  \tilde{L}
  +\tilde{E}^{\dagger} \tilde{m}_{\!\scriptscriptstyle {E}}^2
\, \tilde{E} + \tilde{m}_{\!\scriptscriptstyle
H_{\!\scriptscriptstyle u}}^2 \, |H_{u}|^2
\nonumber \\
&& + \frac{M_{\!\scriptscriptstyle 1}}{2} \tilde{B}\tilde{B}
   + \frac{M_{\!\scriptscriptstyle 2}}{2} \tilde{W}\tilde{W}
   + \frac{M_{\!\scriptscriptstyle 3}}{2} \tilde{g}\tilde{g}
+ {\rm h.c.} \; , \label{soft} \eeqa where we have used ${H}_{d}$
in the place of the equivalent $\tilde{L}_0$ among the trilinear
$A$ terms. Note that $\tilde{L}^\dagger
\tilde{m}_{\!\scriptscriptstyle \tilde{L}}^2  \tilde{L}$, unlike
the other soft mass terms, is given by a $4\times 4$ matrix.
Compared to the MSSM case, $\tilde{m}_{\!\scriptscriptstyle
{L}_{00}}^2$ corresponds to $\tilde{m}_{\!\scriptscriptstyle
H_{\!\scriptscriptstyle d}}^2$ while
$\tilde{m}_{\!\scriptscriptstyle {L}_{0k}}^2$'s give new mass
mixings.

%%%%%%%%%%% Section 3 %%%%%%%%%%%%%%%%%%%%%%%%%%%%%%%%%%%%%%%%%%%%%%%%
\section{Leptonic Radiative Decays}
%%%%%%%%%%%%%%%%%%%%%%%%%%%%%%%%%%%%%%%%%%%%%%%%
Within the GSSM, the three SM charged leptons are the light mass
eigenstates out of a $5\times 5$ charged fermions mass matrix,
which also includes the charginos. We use the common notation
$\chi_{n}^{\scriptscriptstyle \pm}$, $n=1$ to 5, with the former
states given by $\ell_i^{\scriptscriptstyle \pm} \equiv
\chi_{i+2}^{\scriptscriptstyle \pm}$, $i=1$ to 3. The states have
characters different from the fermionic components $l_i^{\!\!\mbox{
-}}$'s and $l_i^{\scriptscriptstyle +}$'s of the $\hat{L}_i$ and
$\hat{E}_i^{\scriptscriptstyle C}$ superfields, respectively, as a
result of the generally nonzero $\mu_i$ RPV mixings between the
charged leptons and charginos of the R-parity conserving (MSSM)
limit. The smallness of the $\mu_i$ values as indicated by the
resulted neutrino mass value \cite{Bisset:1998bt,nu} was the basis
for most of the approximations on related subject matters in the
literature, essentially neglecting the difference between
$\ell_i$'s and $l_i$'s. Reference \cite{Cheung:2001sb} is not totally
free from the kind of approximation, though it focuses on the
${\mu_i^*}\,{\lambda_{ijk}}$ RPV contributions to $\mu \to e \,
\gamma$. We will see that in this kind of parameter combination
the approximation in Ref.\cite{Cheung:2001sb}, which neglects
Higgsino and wino components of the decaying and product charged leptons,
is perfectly fine. The study is the first of its kind, catching a
major role of the $\mu_i$'s, as well as the soft bilinear RPV
parameters $B_i$'s, in the LFV process in conjunction with the
$\lambda$-type couplings. \footnote{See, however, studies on the
${\mu_i^*}\,{\lambda_{ijk}^{\prime}}$ RPV contributions for
similar processes in the quark sector\cite{edms}. In fact, the
general relevancy of the kind of parameter combinations to
flavor diagonal and off-diagonal dipole moments for fermions was
first pointed out in Ref.\cite{as5}.}

As advertised, our analysis here goes beyond that. Let us start by
looking into the full mass eigenstate couplings of the truly
physical charged leptons.

%======================================================== cs-nf vertices
\subsection{Charged scalar vertices}
%=========================================================================
A charged lepton $\ell_i^{\scriptscriptstyle \pm}$ (or a generic
$\chi_{n}^{\scriptscriptstyle \pm}$) couples to a charged scalar
and a neutral fermion. From carefully expanding the Lagrangian, we
have the vertices
\begin{equation}
\label{int1} %{\cal L} =
 {g_{\scriptscriptstyle 2}} \;
\overline{\Psi}({\chi_{\bar{n}}^{\!\!\mbox{ -}}}) \left[ {\cal
N}^{\scriptscriptstyle  L}_{\scriptscriptstyle \bar{n}mn} \,
 {1 - \gamma_{\scriptscriptstyle 5} \over 2} +
{\cal N}^{\scriptscriptstyle  R}_{\scriptscriptstyle \bar{n}mn} \,
{1+ \gamma_{\scriptscriptstyle 5} \over 2} \right ] \,
{\Psi}({\chi}^{\scriptscriptstyle 0}_n) \, \phi_{m}^{\!\!\mbox{
-}} \; + \mbox{h.c.} \;,
\end{equation}
where ${1 \over 2}(1 \mp \gamma_{\scriptscriptstyle 5})$ are the
$L$- and $R$-handed projections and
\begin{eqnarray}
{\cal N}^{\scriptscriptstyle  R}_{\scriptscriptstyle \bar{n}mn}
&=& \mbox{\boldmath $U$}_{\!1\bar{n}}^* \, \mbox{\boldmath
$X$}_{\!\!4n}^{*} \, {\cal D}^{l}_{2m} +\mbox{\boldmath
$U$}_{\!1\bar{n}}^* \, \mbox{\boldmath $X$}_{\!\!(k+4)n}^{*} \,
{\cal D}^{l}_{(k+2)m} \nonumber \\ &&
 + \frac{y_{\!\scriptscriptstyle e_k}}{g_{\scriptscriptstyle 2}} \,
\mbox{\boldmath $U$}_{\!2\bar{n}}^* \, \mbox{\boldmath
$X$}_{\!\!(k+4)n}^{*} \, {\cal D}^{l}_{(k+5)m} +
\frac{1}{\sqrt{2}} \, \mbox{\boldmath $U$}_{\!2\bar{n}}^* \, [
\tan\!\theta_{\!\scriptscriptstyle W} \, \mbox{\boldmath
$X$}_{\!\!1n}^{*} + \mbox{\boldmath $X$}_{\!\!2n}^{*} ] \, {\cal
D}^{l}_{2m} \nonumber \\ && +\frac{1}{\sqrt{2}} \, \mbox{\boldmath
$U$}_{\!(j+2)\bar{n}}^* \, [ \tan\!\theta_{\!\scriptscriptstyle W}
\, \mbox{\boldmath $X$}_{\!\!1n}^{*} + \mbox{\boldmath
$X$}_{\!\!2n}^{*} ] \, {\cal D}^{l}_{(j+2)m}
 - \frac{y_{\!\scriptscriptstyle e_j}}{g_{\scriptscriptstyle 2}} \,
\mbox{\boldmath $U$}_{\!(j+2)\bar{n}}^* \, \mbox{\boldmath
$X$}_{\!\!4n}^{*} \, {\cal D}^{l}_{(j+5)m} \nonumber \\ && -
{\lambda_{kjh}^{\!*} \over g_{\scriptscriptstyle 2} }\,\,
\mbox{\boldmath $U$}_{\!(j+2)\bar{n}}^* \, \mbox{\boldmath
$X$}_{\!(k+4)n}^{*} \, {\cal D}^{l}_{\!(h+5)m}  \;,
\label{Nrnmn} \\
{\cal N}^{\scriptscriptstyle  L}_{\scriptscriptstyle \bar{n}mn}
&=& -\mbox{\boldmath $V$}_{\!1\bar{n}}^* \, \mbox{\boldmath
$X$}_{\!\!4n}^*  \, {\cal D}^{l}_{2m} -\mbox{\boldmath
$V$}_{\!1\bar{n}}^* \, \mbox{\boldmath $X$}_{\!\!(k+4)n}^*  \,
{\cal D}^{l}_{(k+2)m} \nonumber \\ && + \frac{1}{\sqrt{2}} \,
\mbox{\boldmath $V$}_{\!2\bar{n}}^* \, [
-\tan\!\theta_{\!\scriptscriptstyle W} \, \mbox{\boldmath
$X$}_{\!\!1n} + \mbox{\boldmath $X$}_{\!\!2n} ] \, {\cal
D}^{l}_{1m} \nonumber \\ && - \sqrt{2}\,
\tan\!\theta_{\!\scriptscriptstyle W} \mbox{\boldmath
$V$}_{\!\!(j+2)\bar{n}}^* \, \mbox{\boldmath $X$}_{\!\!1n} \,
 {\cal D}^{l}_{(j+5)m}
- \frac{y_{\!\scriptscriptstyle e_j}}{g_{\scriptscriptstyle 2}} \,
\mbox{\boldmath $V$}_{\!\!(j+2)\bar{n}}^* \, \mbox{\boldmath
$X$}_{\!\!4n} {\cal D}^{l}_{(j+2)m} \nonumber \\ && +
\frac{y_{\!\scriptscriptstyle e_j}}{g_{\scriptscriptstyle 2}} \,
\mbox{\boldmath $V$}_{\!\!(j+2)\bar{n}}^* \, \mbox{\boldmath
$X$}_{\!(j+4)n} \, {\cal D}^l_{2m} - {\lambda_{khj} \over
g_{\scriptscriptstyle 2} }\,\, \mbox{\boldmath
$V$}_{\!\!(j+2)\bar{n}}^* \, \mbox{\boldmath $X$}_{\!(k+4)n} \,
{\cal D}^{l}_{\!(h+2)m}  \; , \label{Nlnmn}
\end{eqnarray}
with $\bar{n}$ runs from 1 to 5, $n$ from 1 to 7, and $m$ from 1 to
8. {\boldmath $V$} and {\boldmath $U$} are unitary matrices used
to diagonalize the charged fermion mass matrix, and {\boldmath
$X$} is the diagonalizing matrix of the neutral fermion mass
matrix. ${\cal D}^{l}$ is the diagonalizing matrix of the charged
scalar mass matrix \cite{Kong:2002hb}. We quote the corresponding
terms ${\cal N}^{\scriptscriptstyle R}_{\scriptscriptstyle inm}$
and ${\cal N}^{\scriptscriptstyle L}_{\scriptscriptstyle inm}$
from an earlier formula in Ref.\cite{Cheung:2001sb} for comparison:
\begin{eqnarray}
{\cal N}^{\scriptscriptstyle R}_{\scriptscriptstyle inm} &=&
\frac{1}{\sqrt{2}} \, [\tan\!\theta_{\!\scriptscriptstyle W}
\mbox{\boldmath $X$}_{\!\!1n}^{*} + \mbox{\boldmath
$X$}_{\!\!2n}^{*} ] \, {\cal D}^{l}_{(i+2)m}
 - \frac{y_{\!\scriptscriptstyle e_i}}{g_{\scriptscriptstyle 2}} \,
\mbox{\boldmath $X$}_{\!\!4n}^{*} \, {\cal D}^{l}_{(i+5)m} -
{\lambda_{kih}^* \over g_{\scriptscriptstyle 2} }\,\,
\mbox{\boldmath $X$}_{\!\!(k+4)n}^{*} \, {\cal D}^{l}_{\!(h+5)m}
\;,
\nonumber \\
{\cal N}^{\scriptscriptstyle L}_{\scriptscriptstyle inm} &=&
-\sqrt{2}\, \tan\!\theta_{\!\scriptscriptstyle W} \mbox{\boldmath
$X$}_{\!\!1n} \,
 {\cal D}^{l}_{(i+5)m}
- \frac{y_{\!\scriptscriptstyle e_i}}{g_{\scriptscriptstyle 2}} \,
\mbox{\boldmath $X$}_{\!\!4n} {\cal D}^{l}_{(i+2)m}\nonumber \\
&&+\frac{y_{\!\scriptscriptstyle e_i}}{g_{\scriptscriptstyle 2}}
\,
 \mbox{\boldmath $X$}_{\!\!(k+4)n} \, {\cal D}^l_{2m}
- {\lambda_{khi} \over g_{\scriptscriptstyle 2} }\,\,
\mbox{\boldmath $X$}_{\!\!(k+4)n} \, {\cal D}^{l}_{\!(h+2)m}  \;.
\end{eqnarray}
One-loop diagrams formed by the pair of coupling
vertices give rise to a class of contributions to the radiative
decays we call neutralinolike, which obviously does include the
ones with the physical neutralinos among the fermions
running inside the loop. The new contributions come from the first
five terms of Eq.(\ref{Nrnmn}) and first four terms of
Eq.(\ref{Nlnmn}), which are not seen in Ref.\cite{Cheung:2001sb}.
These new terms involve higssinos and winos on the external legs
and are easy to understand. For example, in ${\cal
N}^{\scriptscriptstyle R}_{\scriptscriptstyle \bar{n}mn}$, the
first term denotes the interaction with wino, neutral Higgsino,
and charged Higgs. The second term is the interaction with wino,
$L$-handed sleptons and the neutrino flavor states, while the
third term describes the interaction with charged Higgsino,
neutrino, and $R$-handed sleptons. The last two terms denote the
interactions with charged Higgsino, charged Higgs, and the bino
and wino, respectively. The nonzero $\mu$'s do give the physical
charged leptons some Higgsino or gaugino components.

%==================================================== begin cf
\subsection{Neutral scalar vertices}
%==========================================================
Next, we come to the charginolike contributions. Here we have to
pair up neutral scalar vertices:
\begin{equation}
\label{int2} %{\cal L} =
{g_{\scriptscriptstyle 2}} \;
\overline{\Psi}({\chi_{\bar{n}}^{\!\!\mbox{ -}}}) \left[ {\cal
C}^{\scriptscriptstyle  L}_{\scriptscriptstyle \bar{n}mn} \, {1 -
\gamma_{\scriptscriptstyle 5} \over 2}  + {\cal
C}^{\scriptscriptstyle  R}_{\scriptscriptstyle \bar{n}mn} \, {1 +
\gamma_{\scriptscriptstyle 5} \over 2}  \right] \,
{\Psi}({\chi}^{\!\!\mbox{ -}}_n) \; \phi_{m}^{\scriptscriptstyle
0} \; + \mbox{h.c.} \;,
\end{equation}
where
\begin{eqnarray} \small
 {\cal C}^{\scriptscriptstyle R}_{\scriptscriptstyle \bar{n}mn}
& =&
 \mbox{\boldmath $U$}_{\!1\bar{n}}^*  \, \mbox{\boldmath $V$}_{\!\!2n} \,
{1 \over \sqrt{2}} \, [ {\cal D}^{s}_{\!1m} - i \, {\cal
D}^{s}_{\!6m} ] -  \mbox{\boldmath $U$}_{\!2\bar{n}}^*  \,
\mbox{\boldmath $V$}_{\!\!1n} \, {1 \over \sqrt{2}} \, [ {\cal
D}^{s}_{\!2m} + i \, {\cal D}^{s}_{\!7m} ] \nonumber \\
&& - \mbox{\boldmath $U$}_{\!(j+2)\bar{n}}^* \, \mbox{\boldmath
$V$}_{\!\!1n} \, {1 \over \sqrt{2}} \, [ {\cal D}^{s}_{\!(j+2)m} +
i \, {\cal
D}^{s}_{\!(j+7)m} ] \nonumber \\
&& + \frac{y_{\!\scriptscriptstyle e_j}}{g_{\scriptscriptstyle 2}}
\, \mbox{\boldmath $U$}_{\!2\bar{n}}^* \, \mbox{\boldmath
$V$}_{\!\!(j+2)n} \,   {1 \over \sqrt{2}} \, [ {\cal
D}^{s*}_{\!(j+2)m} - i \, {\cal D}^{s*}_{\!(j+7)m} ] -
\frac{y_{\!\scriptscriptstyle e_j}}{g_{\scriptscriptstyle 2}} \,
 \mbox{\boldmath $U$}_{\!(j+2)\bar{n}}^* \,
\mbox{\boldmath $V$}_{\!\!(j+2)n}  \,  {1 \over \sqrt{2}} \, [
{\cal
D}^{s*}_{\!2m} - i \, {\cal D}^{s*}_{\!7m} ] \nonumber \\
&& -\frac{\lambda_{hjk}^{\!*}}{g_{\scriptscriptstyle 2}} \,
\mbox{\boldmath $U$}_{\!(j+2)\bar{n}}^* \, \mbox{\boldmath
$V$}_{\!\!(k+2)n} \,   {1 \over \sqrt{2}} \, [ {\cal
D}^{s*}_{\!(h+2)m} - i \, {\cal D}^{s*}_{\!(h+7)m} ] \; ,
\label{Crnmn} \\
 {\cal C}^{\scriptscriptstyle L}_{\scriptscriptstyle \bar{n}mn}
& = &
 \mbox{\boldmath $V$}_{\!\!2\bar{n}}^* \,
\mbox{\boldmath $U$}_{\!1n} \,  {1 \over \sqrt{2}} \, [ {\cal
D}^{s*}_{\!1m} + i \, {\cal D}^{s*}_{\!6m} ] -  \mbox{\boldmath
$V$}_{\!1\bar{n}}^*  \, \mbox{\boldmath $U$}_{\!\!2n} \, {1 \over
\sqrt{2}} \, [ {\cal D}^{s*}_{\!2m} - i \, {\cal D}^{s*}_{\!7m} ]
\nonumber \\
&& -  \mbox{\boldmath $V$}_{\!1\bar{n}}^*  \, \mbox{\boldmath
$U$}_{\!\!(j+2)n} \, {1 \over \sqrt{2}} \, [ {\cal
D}^{s*}_{\!(j+2)m} - i \, {\cal D}^{s*}_{\!(j+7)m} ] \nonumber\\
&& + \frac{y_{\!\scriptscriptstyle e_j}}{g_{\scriptscriptstyle 2}}
\,
  \mbox{\boldmath $V$}_{\!\!(j+2)\bar{n}}^* \, \mbox{\boldmath $U$}_{\!2n} \,
{1 \over \sqrt{2}} \, [ {\cal D}^{s}_{\!(j+2)m} + i \, {\cal
D}^s_{\!(j+7)m} ] - \frac{y_{\!\scriptscriptstyle
e_j}}{g_{\scriptscriptstyle 2}} \mbox{\boldmath
$V$}_{\!\!(j+2)\bar{n}}^* \, \mbox{\boldmath $U$}_{\!(j+2)n} \,
{1 \over \sqrt{2}} \, [ {\cal D}^s_{\!2m} + i \, {\cal D}^s_{\!7m}
]
\nonumber \\
&& + {\lambda_{khj} \over g_{\scriptscriptstyle 2} } \,
\mbox{\boldmath $V$}_{\!\!(j+2)\bar{n}}^* \, \mbox{\boldmath
$U$}_{\!(k+2)n} \,  {1 \over \sqrt{2}} \,
 [ {\cal D}^{s}_{\!(h+2)m} + i \, {\cal D}^s_{\!(h+7)m} ] \; .
\label{Clnmn}
\end{eqnarray}
with $n$ and $\bar{n}$ run from 1 to 5 and $m$ from 1 to 10.
${\cal D}^{s}$ is the diagonalizing matrix of the neutral scalar
mass matrix \cite{Kong:2002hb}. Note that $ {\cal
C}^{\scriptscriptstyle L}_{\bar{n}mn}$ is equal to  ${\cal
C}^{\scriptscriptstyle R^*}_{nm\bar{n}}$ by definition. {These are
to replace ${\cal C}_{inm}^{\scriptscriptstyle L}$ and
 ${\cal C}_{imn}^{\scriptscriptstyle R^*}$ of Ref.\cite{Cheung:2001sb}.
\footnote{ %
Recall that in the latter case, one distinguishes the ``external"
charged lepton, approximated by an $l^{\scriptscriptstyle \pm}$,
from an ``internal'' charged fermion mass eigenstate.} %
\small
\begin{eqnarray}
 {\cal C}^{\scriptscriptstyle R}_{\scriptscriptstyle inm}
& =&  - \mbox{\boldmath $V$}_{\!\!1n} \, {1 \over \sqrt{2}} \, [
{\cal D}^{s}_{\!(i+2)m}+ i \, {\cal D}^{s}_{\!(i+7)m} ]
 -
\frac{y_{\!\scriptscriptstyle e_i}}{g_{\scriptscriptstyle 2}} \,
 \mbox{\boldmath $V$}_{\!\!(i+2)n}  \, {1 \over \sqrt{2}} \,
[ {\cal D}^s_{\!2m} - i \, {\cal D}^s_{\!7m} ]\nonumber \\
 &&-
\frac{\lambda_{hik}^*}{g_{\scriptscriptstyle 2}} \,
\mbox{\boldmath $V$}_{\!\!(k+2)n} \, {1 \over \sqrt{2}} \, [ {\cal
D}^s_{\!(h+2)m} - i \, {\cal D}^s_{\!(h+7)m} ] \; ,
\nonumber \\
 {\cal C}^{\scriptscriptstyle L}_{\scriptscriptstyle inm}
& = &   \frac{y_{\!\scriptscriptstyle e_i}}{g_{\scriptscriptstyle
2}} \,
  \mbox{\boldmath $U$}_{\!2n} \, {1 \over \sqrt{2}} \,
[ {\cal D}^{s}_{\!(i+2)m} + i \, {\cal D}^s_{\!(i+7)m} ] -
\frac{y_{\!\scriptscriptstyle e_i}}{g_{\scriptscriptstyle 2}} \,
\mbox{\boldmath $U$}_{\!(j+2)n} \, {1 \over \sqrt{2}} \, [ {\cal
D}^s_{\!2m} + i \, {\cal D}^s_{\!7m} ]\nonumber \\
&&+ {\lambda_{khi} \over g_{\scriptscriptstyle 2} } \,
\mbox{\boldmath $U$}_{\!(k+2)n} \, {1 \over \sqrt{2}} \, [ {\cal
D}^{s}_{\!(h+2)m} + i \, {\cal D}^s_{\!(h+7)m} ] \; .
\end{eqnarray}
\normalsize %
Note that ${\cal D}^{s}$ is actually real, though we are using
${\cal D}^{s*}$ notation as if it is not. This is just a
convention for tracing the LFV structure of the various
contributions in our analytical discussions below. Here, in fact,
the real difference between the ${\cal D}^{s*}$ and ${\cal D}^{s}$
terms is given explicitly by the different signs between the
corresponding scalar and pseudoscalar parts. Note that the
${y_{\!\scriptscriptstyle e_i}}$ terms in the above expressions
can be written together with the $\lambda$ terms using the
$\lambda_{{\scriptscriptstyle \alpha \beta}k}$ notation and the
identification of ${y_{\!\scriptscriptstyle e_i}}$ as
$\lambda_{{\scriptscriptstyle 0}ii}$. This common structure
between $\hat{L}_0$ and the $\hat{L}_i$'s is very useful in our
discussions below.

%=========================================================================
\subsection{The decay amplitude}
%=========================================================================
In applying the above interactions to the process
$\ell^{\!\!\mbox{ -}}_j (p) \to \ell^{\!\!\mbox{ -}}_i \, \gamma
(q)$, we can write the amplitude as
\begin{equation}
T = e \; \epsilon^{*\scriptscriptstyle \alpha} \, \bar{u}_i (p-q)
\left[ m_{\scriptscriptstyle \ell_{\!j}} \, i \,\sigma_{\!\alpha
\beta} \, q^\beta \left( A_2^{\!\scriptscriptstyle L}  \,
\frac{1-\gamma_{\scriptscriptstyle 5}}{2} +
A_2^{\!\scriptscriptstyle R} \, \frac{1+\gamma_{\scriptscriptstyle
5}}{2} \right )  \right ] u_j(p) \; ,
\end{equation}
where $\epsilon^*=\epsilon^*(q)$ is the polarization four-vector
of the outgoing photon. The decay rate is then simply given by
\begin{equation}
\Gamma({\ell}_{\!\scriptscriptstyle j}^{\!\!\mbox{ -}} \to
{\ell}_{\!\scriptscriptstyle i}^{\!\!\mbox{ -}} \, \gamma) =
{\alpha_{\mbox{\tiny em}} \over 4 } \, m_{\scriptscriptstyle
\ell_{\!j}}^{5} \, \left( \, |A_2^{\!\scriptscriptstyle
L}|^2+|A_2^{\!\scriptscriptstyle R}|^2 \, \right) \; .
\end{equation}
It is straightforward to calculate the contributions from one-loop
diagrams with the effective interactions of Eqs.(\ref{int1}) and
(\ref{int2}).  The result for
 $A_2^{\!\scriptscriptstyle L}$
($A_2^{\!\scriptscriptstyle R}=A_2^{\!\scriptscriptstyle L}|_{L
\leftrightarrow R}$) is given by
%%%%%%%%%%%%%%%%%%%%%%%%
\begin{eqnarray}
A_2^{\!\scriptscriptstyle L}&=& {\alpha_{\mbox{\tiny em}} \over 8
\pi \, \sin\!^2\theta_{\!\scriptscriptstyle W}} \;
% \frac{g_{\scriptscriptstyle 2}^2 }{32  \, \pi^2} \,
%\sum_{m=1}^7 \sum_{n=1}^7
\frac{1}{M_{\!\scriptscriptstyle \tilde{\ell}_{m}}^2} \Bigg[{\cal
N}_{\!\scriptscriptstyle (i+2)mn}^{\scriptscriptstyle L} \,
 {\cal N}_{\!\scriptscriptstyle (j+2)mn}^{\scriptscriptstyle L^*} \,
 F_2\!\!\left({{M}_{\!\scriptscriptstyle \chi^0_{n}}^2 \over
 M_{\!\scriptscriptstyle\tilde{\ell}_{m}}^2} \right)
%\nonumber \\&&
+ {\cal N}_{\!\scriptscriptstyle (i+2)mn}^{\scriptscriptstyle R}\,
 {\cal N}_{\!\scriptscriptstyle (j+2)mn}^{\scriptscriptstyle R^*} \,
\; \frac{m_{\scriptscriptstyle \ell_{\!i}}}{m_{\scriptscriptstyle
\ell_{\!j}}} \,
 F_2\!\!\left({{M}_{\!\scriptscriptstyle \chi^0_{n}}^2 \over
 M_{\!\scriptscriptstyle \tilde{\ell}_{m}}^2} \right) \nonumber \\
% \right. \nonumber \\ && \left.
&&+  {\cal N}_{\!\scriptscriptstyle (i+2)mn}^{\scriptscriptstyle
L} \, {\cal N}_{\!\scriptscriptstyle (j+2)mn}^{\scriptscriptstyle
R^*} \,
 \, {{M}_{\!\scriptscriptstyle \chi^0_{n}}
 \over m_{\scriptscriptstyle \ell_{\!j}}} \,
F_3\!\!\left({{M}_{\!\scriptscriptstyle \chi^0_{n}}^2 \over
M_{\!\scriptscriptstyle \tilde{\ell}_{m}}^2} \right) \Bigg]
%\nonumber \\
%%%%%%%%%%%%%%%%%%%%%%%%%%
%&&
- {\alpha_{\mbox{\tiny em}} \over 8 \pi \,
 \sin\!^2\theta_{\!\scriptscriptstyle W}} \;
% -\frac{g_{\scriptscriptstyle 2}^2}{32 \, \pi^2} \,
\frac{1}{M_{\!\scriptscriptstyle S_{m}}^2} \,
 \Bigg[
 {\cal C}_{\!\scriptscriptstyle (i+2)mn}^{\scriptscriptstyle L} \,
{\cal C}_{\!\scriptscriptstyle (j+2)mn}^{\scriptscriptstyle L^*}
\, F_5\!\!\left({{M}_{\!\scriptscriptstyle \chi^{\mbox{-}}_{n}}^2
\over
M_{\!\scriptscriptstyle S_{m}}^2} \right)\nonumber \\
&&+  {\cal C}_{\!\scriptscriptstyle (i+2)mn}^{\scriptscriptstyle
R} \, {\cal C}_{\!\scriptscriptstyle (j+2)mn}^{\scriptscriptstyle
R^*} \, \frac{m_{\scriptscriptstyle
\ell_{\!i}}}{m_{\scriptscriptstyle \ell_{\!j}}} \,
F_5\!\!\left({{M}_{\!\scriptscriptstyle \chi^{\mbox{-}}_{n}}^2
\over M_{\!\scriptscriptstyle S_{m}}^2}
\right)
%\nonumber \\&&
+{\cal C}_{\!\scriptscriptstyle (i+2)mn}^{\scriptscriptstyle L}\,
 {\cal C}_{\!\scriptscriptstyle (j+2)mn}^{\scriptscriptstyle R^*} \,
{{M}_{\!\scriptscriptstyle \chi^{\mbox{-}}_{n}} \over
m_{\scriptscriptstyle \ell_{\!j}}} \,
F_6\!\!\left({{M}_{\!\scriptscriptstyle \chi^{\mbox{-}}_{n}}^2
\over M_{\!\scriptscriptstyle S_{m}}^2} \right)
 \Bigg], \label{A2L}
\end{eqnarray}
%%%%%%%%%%%%%%%%%%%%%%%%%%%%%
where
\begin{eqnarray}
F_2(x) &=& \frac{1}{6 \, (1-x)^4} \, (1-6 \, x+3 \, x^2+2 \, x^3-6
\, x^2 \, \ln x) \; ,
\nonumber \\
F_3(x) &=& \frac{1}{(1-x)^3} \, (1-x^2+2 \, x \,\ln x) \; ,
\nonumber \\
F_5(x) &=& \frac{1}{6 \,(1-x)^4} \, (2+3 \, x-6 \, x^2+x^3+6 \, x
\, \ln x) \;,
\nonumber \\
F_6(x) &=& \frac{1}{(1-x)^3} \, (-3+4 \, x-x^2-2 \,  \ln x) \; ,
\nonumber
\end{eqnarray}
with summations over all physical fermion and scalar mass
eigenstates as represented by the $n$ and $m$ indices assumed.

The processes we discuss here violate lepton flavor while conserving
the lepton number. Before going into the analysis, it is
instructive to introduce the lepton-flavor numbers $L_e$, $L_\mu$,
and $L_\tau$ to the superfields as one does to their corresponding
components in the SM. The RPV parameters bear violations of the
lepton-flavor numbers. It is obvious that in order to have a
contribution to $\ell^{\!\!\mbox{ -}}_j  \to \ell^{\!\!\mbox{
-}}_i \, \gamma$, a term must reduce $L_j$ and increase $L_i$ by
exactly one unit while leaving the rest unchanged. For instance,
${\mu_1^*}\,{\mu_2}$ means increasing a $L_e$ and reducing a
$L_\mu$ while leaving $L_\tau$ unchanged. This simple but useful
rule serves as a countercheck of individual contributions
discussed below.

%%%%%%%%%%%%%%%%% Section 4 %%%%%%%%%%%%%%%%%%%%%%%%%%%%%%%%%%%%%%%%%%
\section{Numerical Results and Discussions}
%%%%%%%%%%%%%%%%%%%%%%%%%%%%%%%%%%%%%%%%%%%%%%%%%%%%%%%%%%%%%%%%%%%%%%
In this section, we present the results we obtained by a careful
numerical implementation of our $\mu\to e \,\gamma$, $\tau \to
e\,\gamma$, and $\tau \to\mu \,\gamma$ formula with explicit
numerical diagonalization of all the mass matrices involved. We
isolate various major contributions by singling out each of the
corresponding RPV parameter combinations as only nonvanishing
one at a time.  The soft SUSY breaking contributions to R-parity
conserving slepton mixings are set to zero (~{\it i.e.}
$\tilde{m}^2_{\!\scriptscriptstyle L}$,
$\tilde{m}^2_{\!\scriptscriptstyle E}$, and
$A^{\!\scriptscriptstyle E}$ are set to be diagonal~). A basic set
of typical values chosen for the input parameters are given in
Table~\ref{table1}. We used this set of inputs unless otherwise
specified in the results below. Summary of bounds on various
combinations of two R-parity violating parameters is shown in Table~\ref{table2}. 
Neutrino mass bounds are also put onto
the plots. We perform numerical calculation of the neutrino mass
contributions from various parameters involved based on formulas from
Ref.\cite{Kang:2002nq}. The bounds are obtained naively by requiring that
individual neutrino mass terms obtained be less than the sub-eV level.
Note that each of these bounds has different dependence on the background model parameters and are also typically different from the leptonic decay 
contribution term plotted. Hence, the comparison has only a simple
illustrative value. A more comprehensive cross analysis is, however,
considered not appropriate without taking further assumption on some
of the parameters involved. At the end, we also show some of
the effects of varying these input parameters.

%=========================================================no new contribution terms
\subsection{\boldmath\protect The $|\mu^*\lambda|$ or $|B^* \lambda|$ contributions}
%=========================================================================
We first look at the contributions with a ($\mu^*\lambda$) or
($\mu \lambda^*$) structure. The dominant terms do not involve the
Higgsino or wino component of the decaying or the product  charged
leptons. They come from the third term of ${\cal
C}^{\scriptscriptstyle R}_{\bar{n}mn}$ and the only one
$\lambda$-coupling vertex of ${\cal C}^{\scriptscriptstyle
L}_{\bar{n}mn}$, corresponding to the diagrams with the chirality
flip on the internal fermion line. Indeed, the role of the
$\mu_i$'s come in through the internal fermion line, as discussed
in good detail in Ref.~\cite{Cheung:2001sb}. Take
$A^{\!\scriptscriptstyle L}_2$, for example, the dominating term
comes from  ${\cal C}^{\scriptscriptstyle L}_{\bar{n}'mn}$\,${\cal
C}^{\scriptscriptstyle R^*}_{\bar{n}mn}$, where $\bar{n}'<
\bar{n}$. We then have the real scalar part of the contribution,
for example, proportional to \beq \label{mulamda1} \sum_{n=1}^5
\sum_{m=1}^{5} \, \mbox{\boldmath
$U$}_{\!\!(j+2)\bar{n}}\,\mbox{\boldmath
$V$}^{\!*}_{\!\!(j'+2)\bar{n}'}\,\mbox{\boldmath
$V$}^{\!*}_{\!\!1n}\, {M}_{\!\scriptscriptstyle \chi^{\mbox{-}}_n}
 \mbox{\boldmath $U$}_{\!(k+2)n} \;
F_6\!\!\left({{M}_{\!\scriptscriptstyle \chi^{\mbox{-}}_{n}}^2
\over M_{\!\scriptscriptstyle S_{m}}^2} \right)  \; {\cal
D}^{s*}_{\!(j+2)m} \, {\cal D}^{s}_{\!(h+2)m} \;
\frac{\lambda_{khj'}}{g_{\scriptscriptstyle 2}}\; \; . \eeq In the
$\mu \to e\, \gamma$ case,  we have $\bar{n}'=3$ and $\bar{n}=4$
and then use the relation $\mbox{\boldmath $U$}_{\!(j+2)\bar{n}}
\sim \,\delta_{(j+2),\bar{n}}$ and $\mbox{\boldmath
$V$}^{\!*}_{\!(j'+2)\bar{n}'} \sim \,\delta_{(j'+2),\bar{n}'}$.
The expression (\ref{mulamda1}) can be given by \beq
\label{mulamda2} \sum_{n=1}^5 \sum_{m=1}^{5} \, \mbox{\boldmath
$V$}^{\!*}_{\!\!1n}\, {M}_{\!\scriptscriptstyle \chi^{\mbox{-}}_n}
 \mbox{\boldmath $U$}_{\!(k+2)n} \;
F_6\!\!\left({{M}_{\!\scriptscriptstyle \chi^{\mbox{-}}_{n}}^2
\over M_{\!\scriptscriptstyle S_{m}}^2} \right)  \; {\cal
D}^{s*}_{\!4m} \, {\cal D}^{s}_{\!(h+2)m} \;
\frac{\lambda_{kh\scriptscriptstyle 1}}{g_{\scriptscriptstyle
2}}\; \; .
 \eeq
If the loop function $F_6$ could be factored out from the double
summation, we would have a $\mbox{\boldmath $V$}^{\!*}_{\!\!1n}\,
{M}_{\!\scriptscriptstyle \chi^{\mbox{-}}_n}
 \mbox{\boldmath $U$}_{\!(k+2)n}$
summation over fermions and a ${\cal D}^{s*}_{\!4m} \, {\cal
D}^{s}_{\!(h+2)m}$ summation over (real) scalars. Taking $h=2$ in
the above expression (\ref{mulamda2}), we have the two dominating
chargino contributions, the $n=1$ and $2$ parts, given
approximately by \beq \label{1st+} \mbox{\boldmath
$V$}^{\!*}_{\!\!1n}\, {\mu_k^*} \, R_{\!\scriptscriptstyle R_{2n}}
\, \frac{\lambda_{k\scriptscriptstyle 21}}{g_{\scriptscriptstyle
2}} \; F_6\!\!\left({{M}_{\!\scriptscriptstyle
\chi^{\mbox{-}}_{n}}^2 \over M_{\!\scriptscriptstyle S_{m}}^2}
\right)  \;, \eeq where $R_{\!\scriptscriptstyle R}$ is a $2\times
2$ matrix with order 1 matrix elements  (see
Ref.\cite{Kong:2002hb} for details.) The expected combination
$\mu_k^*\,\lambda_{k\scriptscriptstyle 21}$ comes up, with $k=1$
and $3$ admissible. The same situation goes for the ${\cal
C}^{\scriptscriptstyle R}_{3mn}\, {\cal C}^{\scriptscriptstyle
L^*}_{4mn}$ part of  $A^{\!\scriptscriptstyle R}_2$, with the
combination $\mu_k \,\lambda_{k\scriptscriptstyle 12}^*$  (~$k=2$
and $3$ admissible here~) instead.

Terms involving the Higgsino and wino components of the decaying
and product charged leptons are expected to be proportional to
$\mbox{\boldmath $U$}_{\!\!a\bar{n}}$
 or $\mbox{\boldmath $V$}_{\!\!a\bar{n}}$ for $a=1$ or 2, with\cite{Kong:2002hb}
\begin{eqnarray} \label{Blam2}
\mbox{\boldmath $U$}_{\!\!a(i+2)} & \varpropto &
 \mu_i\;,
\nonumber \\
\mbox{\boldmath $V$}_{\!\!a(i+2)} & \varpropto &
 \mu_i\, m_i \; .
\end{eqnarray}
In the case of the $\mu \to e\, \gamma$, if we focus on the
$\mu_3^*$ and $\lambda_{\scriptscriptstyle 321}$ parameter
combination, the new contributions of ${\cal
N}^{\scriptscriptstyle L,R}_{\scriptscriptstyle {3}mn}$, ${\cal
N}^{\scriptscriptstyle L,R}_{\scriptscriptstyle {4}mn}$, ${\cal
C}^{\scriptscriptstyle L,R}_{\scriptscriptstyle {3}mn}$, and
${\cal C}^{\scriptscriptstyle L,R}_{\scriptscriptstyle {4}mn}$ do
not contribute to the $Br(\mu \to e\, \gamma)$, since they are
only proportional to the $\mu_1$ or $\mu_2$, not $\mu_3$.
Likewise, the $\mu_3 \,\lambda_{\scriptscriptstyle 312}^*$ also
does not have a new contribution to $Br(\mu \to e\, \gamma)$.
%-------------------------The combination of b-lambda doesn't change ---
For the same reason, the contributions of $\mu_1
\,\lambda_{\scriptscriptstyle 123}^*$, $\mu_1^*
\,\lambda_{\scriptscriptstyle 132}$, $\mu_2
\,\lambda_{\scriptscriptstyle 213}^*$, $\mu_2^*
\,\lambda_{\scriptscriptstyle 231}$, and all ${B^*\, \lambda}$ and
${B \, \lambda^*}$ type combinations to the branching ratio of
leptonic radiative decays are still essentially the same as the
results obtained by neglecting the wino and higgino components of
the decaying and product charged leptons.

There are, however, new contributions that come from  the
charginolike loop diagrams with the chirality flip on the
external fermion line. They are the product of the fourth term and
the $\lambda$-coupling term of ${\cal C}^{\scriptscriptstyle
R}_{\bar{n}mn}$. There are two types of diagrams. The first type
comes from ${\cal C}^{\scriptscriptstyle R}_{\bar{n}'mn}$\,${\cal
C}^{\scriptscriptstyle R^*}_{\bar{n}mn}$, where $\bar{n}'<
\bar{n}$, with a $\lambda$ coupling in the ${\cal
C}^{\scriptscriptstyle R^*}_{\bar{n}mn}$ and a Yukawa coupling in
the ${\cal C}^{\scriptscriptstyle R}_{\bar{n}'mn}$. It is given by
\beq \label{Blam2} m_{\scriptscriptstyle \ell}
\sum_{n=1}^5\sum_{m=1}^{5} \mbox{\boldmath
$U$}^{\!*}_{\!2\bar{n}'}\, \mbox{\boldmath $U$}_{\!(j+2)\bar{n}}
\; \mbox{\boldmath $V$}_{\!(j'+2)n}\,
 \mbox{\boldmath $V$}^{\!*}_{\!(k+2)n} \;
F_5\!\!\left({{M}_{\!\scriptscriptstyle \chi^{\mbox{-}}_{n}}^2
\over M_{\!\scriptscriptstyle S_{m}}^2} \right)  \; {\cal
D}^{s*}_{\!(j'+2)m} \, {\cal D}^{s}_{\!(h+2)m} \; \frac{-
\lambda_{hjk}}{g_{\scriptscriptstyle 2}}\;
\frac{y_{\!\scriptscriptstyle e_{j'}}}{g_{\scriptscriptstyle 2}}\;
, \eeq
requiring further $h=j'$ and $k=j'$. Note that all off-diagonal
matrix elements of the form $\mbox{\boldmath $V$}_{\!\!(k+2)n}$
are very small, those RPV ones ($n=1$ or 2), in particular, contain
a ${y_{\!\scriptscriptstyle e_k}}$ suppression \cite{Kong:2002hb}.
A similar situation goes with the scalar sum ${\cal
D}^{s*}_{\!(j'+2)m} \, {\cal D}^{s}_{\!(h+2)m} = \delta_{j'h}$ by
unitarity.  $m_{\scriptscriptstyle \ell}$ stands for the mass of
the decaying lepton. There is a factor of $y_{\!\scriptscriptstyle
e_{j'}}$ suppression in the expression (\ref{Blam2}). The
electroweak state Feynman diagram of this expression is in Fig.~1.
Note the Higgsino component of an external line illustrated.
%%%%%%%%%%%%%%%%%%%%%%%%%%%%%%% example %%%%%%%

In the case of $\tau \to \mu\, \gamma$, we have $\bar{n}'=4$ and
$\bar{n}=5$. After we take $j'=h=k=2$ in the expression
(\ref{Blam2}) and use the relation $\mbox{\boldmath
$U$}_{\!(j+2)\bar{n}} \sim \,\delta_{(j+2),\bar{n}}$ and
$\mbox{\boldmath $U$}^{*}_{\!\!2(i+2)}  \varpropto  \mu^{*}_i$. ,
the expression (\ref{Blam2}) can be given approximately by \beq
\label{Blam3} m_{\scriptscriptstyle \tau}
\mu^{*}_{\scriptscriptstyle 2} \frac{
\lambda_{232}}{g_{\scriptscriptstyle 2}}\;
\frac{y_{\!\scriptscriptstyle e_{2}}}{g_{\scriptscriptstyle 2}}\;
F_5\!\!\left({{M}_{\!\scriptscriptstyle \chi^{\mbox{-}}_{n}}^2
\over M_{\!\scriptscriptstyle S_{m}}^2} \right)  \;.
\eeq
The expected $\mu_{\scriptscriptstyle 2}^*
\,\lambda_{\scriptscriptstyle 232}$ comes out with a factor of
$y_{\!\scriptscriptstyle e_{2}}$ (muon Yukawa coupling)
suppression. The same situation goes for the case of the $\mu \to
e\,\gamma$ and $\tau \to e\, \gamma$,
corresponding to the combination of %the RPV parameter
 $\mu_{\scriptscriptstyle 1}^* \,\lambda_{\scriptscriptstyle 121}$ and
$\mu_{\scriptscriptstyle 1}^* \,\lambda_{\scriptscriptstyle 131}$,
respectively, with a $y_{\!\scriptscriptstyle e_{1}}$ suppression.

%%%%%%%%%%%%%%%%% Second type %%%%%%%%%%%%%%%%%%%%%%%
Likewise, the second type comes from ${\cal C}^{\scriptscriptstyle
R}_{\bar{n}'mn}$\,${\cal C}^{\scriptscriptstyle R^*}_{\bar{n}mn}$,
where $\bar{n}'< \bar{n}$, with a $\lambda$ coupling in the
${\cal C}^{\scriptscriptstyle R}_{\bar{n}'mn}$ and a Yukawa
coupling in the ${\cal C}^{\scriptscriptstyle R^*}_{\bar{n}mn}$.
It is given by \beq \label{Blam4} m_{\scriptscriptstyle \ell}
\sum_{n=1}^5\sum_{m=1}^{5} \mbox{\boldmath $U$}_{\!2\bar{n}}\,
\mbox{\boldmath $U$}^{\!*}_{\!(j'+2)\bar{n}'} \; \mbox{\boldmath
$V$}^{\!*}_{\!(j+2)n}\,
 \mbox{\boldmath $V$}_{\!(k+2)n} \;
F_5\!\!\left({{M}_{\!\scriptscriptstyle \chi^{\mbox{-}}_{n}}^2
\over M_{\!\scriptscriptstyle S_{m}}^2} \right)  \; {\cal
D}^{s}_{\!(j+2)m} \, {\cal D}^{s*}_{\!(h+2)m} \; \frac{-
\lambda^*_{hj'k}}{g_{\scriptscriptstyle 2}}\;
\frac{y_{\!\scriptscriptstyle e_{j}}}{g_{\scriptscriptstyle 2}}\;
. \eeq The electroweak state Feynman diagram for this case is in
Fig.~\ref{00Chen20}. In the case of $\mu \to e\, \gamma$, we have
$\bar{n}'=3$ and $\bar{n}=4$. After taking $j=h=k=2$ and using the
relation mentioned above, the expected $\mu_{\scriptscriptstyle 2}
\,\lambda_{\scriptscriptstyle 212}^*$ would come out, with a
factor of muon Yukawa. The same situation goes for the case of the
$\tau \to e\,\gamma$, and $\tau \to\mu\, \gamma$, corresponding to
the combination of $\mu_{\scriptscriptstyle 3}
\,\lambda_{\scriptscriptstyle 313}^*$ and $\mu_{\scriptscriptstyle
3} \,\lambda_{\scriptscriptstyle 323}^*$, respectively, with a tau
Yukawa. Compared with the first type, the second one does have
larger contributions, since the former have the stronger Yukawa
suppression. This result is confirmed by our exact numerical
calculation and could be understood easily through
Fig.~\ref{00Chen10} and Fig.~\ref{00Chen20}, where the relevant
illustrative electroweak state one-loop diagrams are given.
Because these two types of contributions involve leptonic Yukawa
couplings, they have $\frac{1}{\cos \beta}$ dependence. However,
they are not the dominant contributions, so the $\mu^*\lambda$ or
$\mu \lambda^*$ type contributions are still insensitive to the
$\tan \!\beta$.

We plot contours of the resulting branching ratio as a function of
(real) ${\mu_{\scriptscriptstyle 2}}$ and
$\lambda_{\scriptscriptstyle 212}$, ${\mu_{\scriptscriptstyle 3}}$
and $\lambda_{\scriptscriptstyle 323}$, and
${\mu_{\scriptscriptstyle 3}}$ and $\lambda_{\scriptscriptstyle
313}$ in Fig.~\ref{ML1}, Fig.~\ref{ML6}, and Fig.~\ref{ML10},
respectively. The present experimental limit is also shown and the
allowed region at 90\% C.L. is shaded. The 90\% C.L. upper limit
on $|{\mu_k^*}\,{\lambda_{k\scriptscriptstyle 21}}|$ or
$|{\mu_k}\,{\lambda_{k\scriptscriptstyle 12}^*}|$ (normalized by
$|\mu_{\scriptscriptstyle 0}|= 135\,\mbox{GeV}$) is given by
\begin{equation}
\frac{|{\mu_{\scriptscriptstyle
3}^*}\,{\lambda_{\scriptscriptstyle 321}}|}
{|\mu_{\scriptscriptstyle 0}|}\;, \;\;\;
\frac{|{\mu_{\scriptscriptstyle
1}^*}\,{\lambda_{\scriptscriptstyle 121}}|}
{|\mu_{\scriptscriptstyle 0}|}\;, \;\;\;
\frac{|{\mu_{\scriptscriptstyle 3}}\,{\lambda_{\scriptscriptstyle
312}^*}|} {|\mu_{\scriptscriptstyle 0}|}\;, \;\;\;
\frac{|{\mu_{\scriptscriptstyle 2}}\,{\lambda_{\scriptscriptstyle
212}^*}|} {|\mu_{\scriptscriptstyle 0}|}\; \;\;\; < 2.1 \times
10^{-7} \;.
\end{equation}
Likewise, the 90\% C.L. upper limit on
$|{\mu_k^*}\,{\lambda_{k\scriptscriptstyle 32}}|$ or
$|{\mu_k}\,{\lambda_{k\scriptscriptstyle 23}^*}|$ (normalized by
$|\mu_{\scriptscriptstyle 0}|= 135\,\mbox{GeV}$) is given by
\begin{equation}
\frac{|{\mu_{\scriptscriptstyle
2}^*}\,{\lambda_{\scriptscriptstyle 232}}|}
{|\mu_{\scriptscriptstyle 0}|}\;, \;\;\;
\frac{|{\mu_{\scriptscriptstyle
1}^*}\,{\lambda_{\scriptscriptstyle 132}}|}
{|\mu_{\scriptscriptstyle 0}|}\;, \;\;\;
\frac{|{\mu_{\scriptscriptstyle 3}}\,{\lambda_{\scriptscriptstyle
323}^*}|} {|\mu_{\scriptscriptstyle 0}|}\;, \;\;\;
\frac{|{\mu_{\scriptscriptstyle 1}}\,{\lambda_{\scriptscriptstyle
123}^*}|} {|\mu_{\scriptscriptstyle 0}|}\; \;\;\; < 7.0 \times
10^{-4} \;.
\end{equation}
The 90\% C.L. upper limit on
$|{\mu_k^*}\,{\lambda_{k\scriptscriptstyle 31}}|$ or
$|{\mu_k}\,{\lambda_{k\scriptscriptstyle 13}^*}|$ (normalized by
$|\mu_{\scriptscriptstyle 0}|= 135\,\mbox{GeV}$) is given by
\begin{equation}
\frac{|{\mu_{\scriptscriptstyle
2}^*}\,{\lambda_{\scriptscriptstyle 231}}|}
{|\mu_{\scriptscriptstyle 0}|}\;, \;\;\;
\frac{|{\mu_{\scriptscriptstyle
1}^*}\,{\lambda_{\scriptscriptstyle 131}}|}
{|\mu_{\scriptscriptstyle 0}|}\;, \;\;\;
\frac{|{\mu_{\scriptscriptstyle 3}}\,{\lambda_{\scriptscriptstyle
313}^*}|} {|\mu_{\scriptscriptstyle 0}|}\;, \;\;\;
\frac{|{\mu_{\scriptscriptstyle 2}}\,{\lambda_{\scriptscriptstyle
213}^*}|} {|\mu_{\scriptscriptstyle 0}|}\; \;\;\; < 8.5 \times
10^{-4}\;.
\end{equation}

For the $B \lambda^*$ structure, one of the contributions is from
the chirality flip inside the loop. In $A_2^{\!\scriptscriptstyle
L}$, this term comes from the $\lambda$ coupling term of ${\cal
C}^{\scriptscriptstyle L}_{\bar{n'}mn}$ and the fifth term of
${\cal C}^{\scriptscriptstyle R}_{\bar{n}mn}$, where $ \bar{n}' <
\bar{n}$. We then have the expression
\beqa \label{Blambda1}
 &&
\sum_{m}^{\prime} \sum_{n=1}^{5}
\,\mbox{\boldmath $V$}^{\!*}_{\!(j'+2)\bar{n}'}\, \mbox{\boldmath
$U$}_{\!(j+2)\bar{n}} \; \mbox{\boldmath $V$}^{\!*}_{\!(j+2)n}\,
{M}_{\!\scriptscriptstyle \chi^{\mbox{-}}_n}
 \mbox{\boldmath $U$}_{\!(k+2)n} \;
F_6\!\!\left({{M}_{\!\scriptscriptstyle \chi^{\mbox{-}}_{n}}^2
\over M_{\!\scriptscriptstyle S_{m}}^2} \right)
\nonumber \\ && \hspace*{1.5in}
\cdot
 [{\cal D}^{s}_{\!2m} + i {\cal D}^{s}_{\!7m} ] \, [{\cal
D}^{s}_{\!(h+2)m} + i {\cal D}^{s}_{\!(h+7)m} ] \;
\frac{y_{\!\scriptscriptstyle e_j}}{g_{\scriptscriptstyle 2}}\;
\frac{-\lambda_{hkj'} }{g_{\scriptscriptstyle 2}}\; , \eeqa
where the $\sum_{m}^{\prime}$ notation means the unphysical Goldstone
mode is omitted. In the case of $\mu \to e\, \gamma$, we have
$\bar{n}'=3$ and $\bar{n}=4$. Using $\mbox{\boldmath
$V$}^{*}_{\!(j'+2)\bar{n}'} \sim \,\delta_{\!(j'+2)\bar{n}'}$ and
$\mbox{\boldmath $U$}_{\!(j+2)\bar{n}} \sim
\delta_{\!(j+2)\bar{n}}$ and taking $k=2$, we then have the expression
\beq \label{Blambda2} \sum_{m}^{\prime} \sum_{n=1}^{5} \,
\mbox{\boldmath $V$}^{\!*}_{\!\!4n}\, {M}_{\!\scriptscriptstyle
\chi^{\mbox{-}}_n}
 \mbox{\boldmath $U$}_{\!4n} \;
F_6\!\!\left({{M}_{\!\scriptscriptstyle \chi^{\mbox{-}}_{n}}^2
\over M_{\!\scriptscriptstyle S_{m}}^2} \right) [{\cal
D}^{s}_{\!2m} + i {\cal D}^{s}_{\!7m} ] \, [{\cal
D}^{s}_{\!(h+2)m} + i {\cal D}^{s}_{\!(h+7)m} ] \;
\frac{y_{\!\scriptscriptstyle e_2}}{g_{\scriptscriptstyle 2}}\;
\frac{-\lambda_{h\scriptscriptstyle 21} }{g_{\scriptscriptstyle
2}}\; . \eeq The fermionic sum suggests a major contribution from
$n=4$, {\it i.e.} the muon itself with the $m_\mu$ dependence.
However, this contribution then has two factors of ``muon Yukawa"
(${y_{\!\scriptscriptstyle e_2}}$) suppression. The scalar sum
gives a contribution proportional to $B_{j'}^* \, \tan\!\beta$.
The expected combination $B_{h}^*\, \lambda_{h\scriptscriptstyle
2\!1}$ comes up with an explicit $\tan \beta$ dependence.
%%%%%%%%%%%%%%%%%%%%%%%%%%%%%%%%
The other contribution having similar strength to the one in
expression (\ref{Blambda1}) discussed above is from the chirality
flip on the external muon line. In $A_2^{\!\scriptscriptstyle L}$,
this term comes from the $\lambda$ coupling term of ${\cal
C}^{\scriptscriptstyle L}_{\bar{n'}mn}$ and the fifth term of
${\cal C}^{\scriptscriptstyle L}_{\bar{n}mn}$, where $ \bar{n}' <
\bar{n}$. We then have the contribution proportional to
 \beq
\label{Blambda3} m_{\scriptscriptstyle \ell}
\sum_{n=1}^5\sum_{m=1}^{5} \mbox{\boldmath $V$}_{\!(j+2)\bar{n}}\,
\mbox{\boldmath $V$}^{\!*}_{\!(j'+2)\bar{n}'} \; \mbox{\boldmath
$U$}^{\!*}_{\!(j+2)n}\,
 \mbox{\boldmath $U$}_{\!(k+2)n} \;
F_5\!\!\left({{M}_{\!\scriptscriptstyle \chi^{\mbox{-}}_{n}}^2
\over M_{\!\scriptscriptstyle S_{m}}^2} \right)  \; {\cal
D}^{s}_{\!(h+2)m} \, {\cal D}^{s*}_{\!2m} \; \frac{-
\lambda_{hkj'}}{g_{\scriptscriptstyle 2}}\;
\frac{-y_{\!\scriptscriptstyle e_{j}}}{g_{\scriptscriptstyle 2}}\;
. \eeq In the case of $\mu \to e\, \gamma$, we have $\bar{n}'=3$
and $\bar{n}=4$. Using $\mbox{\boldmath $V$}_{\!(j+2)\bar{n}} \sim
\,\delta_{(j+2),\bar{n}}$ and $\mbox{\boldmath
$V$}^{*}_{\!(j'+2)\bar{n'}} \sim  \delta_{\!(j'+2)\bar{n}'}$ and
taking $k=2$, we have the expression \beq \label{Blambda4}
m_{\scriptscriptstyle \mu} \sum_{n=1}^5\sum_{m=1}^{5}
 \mbox{\boldmath $U$}^{\!*}_{\!4n}\,
 \mbox{\boldmath $U$}_{\!4n} \;
F_5\!\!\left({{M}_{\!\scriptscriptstyle \chi^{\mbox{-}}_{n}}^2
\over M_{\!\scriptscriptstyle S_{m}}^2} \right)  \; {\cal
D}^{s}_{\!(h+2)m} \, {\cal D}^{s*}_{\!2m} \; \frac{-
\lambda_{h\scriptscriptstyle 2\!1}}{g_{\scriptscriptstyle 2}}\;
\frac{- y_{\!\scriptscriptstyle e_2}}{g_{\scriptscriptstyle 2}}\;
, \eeq where we have $h=3$ or $1$. The scalar sum gives the
dominating contribution proportional to $B_h^*$. The expected
combination $B_{h}^*\, \lambda_{h\scriptscriptstyle 2\!1}$ comes
up.
%%%%%%%%%%%%%%%%%%%%%%%%%
The 90\% C.L. upper limit on
$|{B_k^*}\,{\lambda_{k\scriptscriptstyle 21}}|$ or
$|{B_k}\,{\lambda_{k\scriptscriptstyle 12}^*}|$ (normalized by
$|\mu_{\scriptscriptstyle 0}|= 135\,\mbox{GeV}$) is given by
\begin{equation}
\frac{|{B_{\scriptscriptstyle 3}^*}\,{\lambda_{\scriptscriptstyle
321}}|} {|\mu_{\scriptscriptstyle 0}|^2}\;, \;\;\;
\frac{|{B_{\scriptscriptstyle 1}^*}\,{\lambda_{\scriptscriptstyle
121}}|} {|\mu_{\scriptscriptstyle 0}|^2}\;, \;\;\;
\frac{|{B_{\scriptscriptstyle 3}}\,{\lambda_{\scriptscriptstyle
312}^*}|} {|\mu_{\scriptscriptstyle 0}|^2}\;, \;\;\;
\frac{|{B_{\scriptscriptstyle 2}}\,{\lambda_{\scriptscriptstyle
212}^*}|} {|\mu_{\scriptscriptstyle 0}|^2}\; \;\;\; < 1.3 \times
10^{-4} \;.
\end{equation}
Likewise, the 90\% C.L. upper limit on
$|{B_k^*}\,{\lambda_{k\scriptscriptstyle 32}}|$ or
$|{B_k}\,{\lambda_{k\scriptscriptstyle 23}^*}|$ (normalized by
$|\mu_{\scriptscriptstyle 0}|= 135\,\mbox{GeV}$) is given by
\begin{equation}
\frac{|{B_{\scriptscriptstyle 2}^*}\,{\lambda_{\scriptscriptstyle
232}}|} {|\mu_{\scriptscriptstyle 0}|^2}\;, \;\;\;
\frac{|{B_{\scriptscriptstyle 1}^*}\,{\lambda_{\scriptscriptstyle
132}}|} {|\mu_{\scriptscriptstyle 0}|^2}\;, \;\;\;
\frac{|{B_{\scriptscriptstyle 3}}\,{\lambda_{\scriptscriptstyle
323}^*}|} {|\mu_{\scriptscriptstyle 0}|^2}\;, \;\;\;
\frac{|{B_{\scriptscriptstyle 1}}\,{\lambda_{\scriptscriptstyle
123}^*}|} {|\mu_{\scriptscriptstyle 0}|^2}\; \;\;\; < 1.4 \times
10^{-3} \;.
\end{equation}
The 90\% C.L. upper limit on
$|{B_k^*}\,{\lambda_{k\scriptscriptstyle 31}}|$ or
$|{B_k}\,{\lambda_{k\scriptscriptstyle 13}^*}|$ (normalized by
$|\mu_{\scriptscriptstyle 0}|= 135\,\mbox{GeV}$) is given by
\begin{equation}
\frac{|{B_{\scriptscriptstyle 2}^*}\,{\lambda_{\scriptscriptstyle
231}}|} {|\mu_{\scriptscriptstyle 0}|^2}\;, \;\;\;
\frac{|{B_{\scriptscriptstyle 1}^*}\,{\lambda_{\scriptscriptstyle
131}}|} {|\mu_{\scriptscriptstyle 0}|^2}\;, \;\;\;
\frac{|{B_{\scriptscriptstyle 3}}\,{\lambda_{\scriptscriptstyle
313}^*}|} {|\mu_{\scriptscriptstyle 0}|^2}\;, \;\;\;
\frac{|{B_{\scriptscriptstyle 2}}\,{\lambda_{\scriptscriptstyle
213}^*}|} {|\mu_{\scriptscriptstyle 0}|^2}\; \;\;\; < 1.9 \times
10^{-3}\;.
\end{equation}

%%=========================================================================
\subsection{\boldmath\protect  The ${|B^*\mu|}$  contributions}
%%=========================================================================
Next, we will discuss the  ${B^*\, \mu}$ and ${B \, \mu^*}$ type
combinations. The dominant terms come from two types of diagrams
with the chirality flip inside the loop. One of them comes from
the third term of ${\cal C}^{\scriptscriptstyle R}_{\bar{n}mn}$
and the fifth term of ${\cal C}^{\scriptscriptstyle
L}_{\bar{n}mn}$. In $A^{\!\scriptscriptstyle R}_2$, this term
comes from ${\cal C}^{\scriptscriptstyle R}_{\bar{n}'mn}$\,${\cal
C}^{\scriptscriptstyle L^*}_{\bar{n}mn}$, where $ \bar{n}' <
\bar{n}$. We then have the real scalar part of the contribution
proportional to \beq \label{2murl2g1} \sum_{n=1}^5\sum_{m=1}^{5}
\mbox{\boldmath $U$}^{\!*}_{\!(j'+2)\bar{n}'}\, \mbox{\boldmath
$V$}_{\!(j+2)\bar{n}} \; \mbox{\boldmath $V$}_{\!1n}\,
{M}_{\!\scriptscriptstyle \chi^{\mbox{-}}_n}
 \mbox{\boldmath $U$}^{\!*}_{\!(j+2)n} \;
F_6\!\!\left({{M}_{\!\scriptscriptstyle \chi^{\mbox{-}}_{n}}^2
\over M_{\!\scriptscriptstyle S_{m}}^2} \right)  \; {\cal
D}^{s}_{\!(j'+2)m} \, {\cal D}^{s*}_{\!2m} \;
\frac{y_{\!\scriptscriptstyle e_{j}}}{g_{\scriptscriptstyle 2}}\;
. \eeq The relevant electroweak state Feynman diagram is then in
Fig.~\ref{00Chen30}. With only $l_i^\pm$ on the external legs, it
obviously involves no Higgsino or wino component there. The
dominating part with the charginos ($n=1$ and $2$) gives a $\mu_j$
dependence through $\mbox{\boldmath $U$}^{\!*}_{\!(j+2)n}$. The
scalar sum gives the contribution proportional to $B_{j'}^* \,
\tan\!\beta$. The expected combination $B_{j'}^*\, \mu_j$ comes up
with an explicit $\tan \beta$ dependence. Similarly, in
$A^{\!\scriptscriptstyle L}_2$, this term comes from ${\cal
C}^{\scriptscriptstyle L}_{\bar{n}'mn}$\,${\cal
C}^{\scriptscriptstyle R^*}_{\bar{n}mn}$, again $ \bar{n}' <
\bar{n}$. It is given by \beq \label{2murl2g2}
\sum_{n=1}^5\sum_{m=1}^{5} \mbox{\boldmath
$V$}^{\!*}_{\!(j'+2)\bar{n}'} \, \mbox{\boldmath
$U$}_{\!(j+2)\bar{n}}\; \mbox{\boldmath $U$}_{\!(j'+2)n}\,
{M}_{\!\scriptscriptstyle \chi^{\mbox{-}}_n} \mbox{\boldmath
$V$}^{\!*}_{\!1n}  \; F_6\!\!\left({{M}_{\!\scriptscriptstyle
\chi^{\mbox{-}}_{n}}^2 \over M_{\!\scriptscriptstyle S_{m}}^2}
\right)  \; {\cal D}^{s}_{\!2m}\, {\cal D}^{s*}_{\!(j+2)m}  \;
\frac{y_{\!\scriptscriptstyle e_{j'}}}{g_{\scriptscriptstyle 2}}\;
. \eeq The expected combination $\mu_{j'}^*\, B_j$ comes up (see
Fig.~\ref{00Chen42}).

%%%%%%%%%%%%%%%%%% 2nd type in mu b %%%%%%%%%%%%%%%%%%%%%%%%%%%%%%%%%%%%%%%

The other type of contribution comes from the second term of ${\cal
C}^{\scriptscriptstyle R}_{\bar{n}mn}$ and the fourth term of
${\cal C}^{\scriptscriptstyle L}_{\bar{n}mn}$, which is a new
contribution not considered in Ref.~\cite{Cheung:2001sb}. It
does have a sizable contribution in some RPV parameter space
regions. In $A^{\!\scriptscriptstyle R}_2$, this term comes from
${\cal C}^{\scriptscriptstyle R}_{\bar{n}'mn}$\,${\cal
C}^{\scriptscriptstyle L^*}_{\bar{n}mn}$. We then have the
contribution proportional to \beq \label{2murl2g3}
\sum_{n=1}^5\sum_{m=1}^{5}- \mbox{\boldmath
$U$}^{\!*}_{\!2\bar{n}'}\, \mbox{\boldmath $V$}_{\!(j+2)\bar{n}}
\; \mbox{\boldmath $V$}_{\!1n}\, {M}_{\!\scriptscriptstyle
\chi^{\mbox{-}}_n}
 \mbox{\boldmath $U$}^{\!*}_{\!2n} \;
F_6\!\!\left({{M}_{\!\scriptscriptstyle \chi^{\mbox{-}}_{n}}^2
\over M_{\!\scriptscriptstyle S_{m}}^2} \right)  \; {\cal
D}^{s}_{\!2m} \, {\cal D}^{s*}_{\!(j+2)m} \;
\frac{y_{\!\scriptscriptstyle e_{j}}}{g_{\scriptscriptstyle 2}}\;
. \eeq We expect a $B_j \,\tan\!\beta$ from the scalar mixing
part, and a $\mu_{\scriptscriptstyle {\bar{n}'-2}}^*$ from
$\mbox{\boldmath $U$}^{\!*}_{\!2\bar{n}'}$. In
$A^{\!\scriptscriptstyle L}_2$, this term comes from ${\cal
C}^{\scriptscriptstyle L}_{\bar{n}'mn}$\,${\cal
C}^{\scriptscriptstyle R^*}_{\bar{n}mn}$. It is given by \beq
\label{2murl2g4} \sum_{n=1}^5\sum_{m=1}^{5}- \mbox{\boldmath
$V$}^{\!*}_{\!(j'+2)\bar{n}'} \, \mbox{\boldmath
$U$}_{\!2\bar{n}}\; \mbox{\boldmath $U$}_{\!2n}\,
{M}_{\!\scriptscriptstyle \chi^{\mbox{-}}_n} \mbox{\boldmath
$V$}^{\!*}_{\!1n}  \; F_6\!\!\left({{M}_{\!\scriptscriptstyle
\chi^{\mbox{-}}_{n}}^2 \over M_{\!\scriptscriptstyle S_{m}}^2}
\right)  \; {\cal D}^{s}_{\!(j'+2)m}\, {\cal D}^{s*}_{\!2m}  \;
\frac{y_{\!\scriptscriptstyle e_{j'}}}{g_{\scriptscriptstyle 2}}\;
. \eeq Again, we expect a $B_{j'}^* \, \tan\!\beta$ from the
scalar mixing part, and a $\mu_{\scriptscriptstyle {\bar{n}-2}}$
from the $\mbox{\boldmath $U$}_{\!2\bar{n}}.$ The relevant
electroweak state Feynman diagrams for the cases are given in
Figs.~\ref{00Chen50} and \ref{00Chen60}, respectively. They clearly
illustrate the role of the Higgsino component on an external line.

%%%%%%%%%%%%%%%%%%%%%% example mu b %%%%%%%%%%%%%%%%%%%%%%%%%%

Taking the $\mu \to e\, \gamma$ for example, we have $\bar{n}'=3$
and $\bar{n}=4$. The expression (\ref{2murl2g1}) would become \beq
\label{2murl1} \sum_{n=1}^5 \sum_{m=1}^{5}  \mbox{\boldmath
$V$}_{\!\!1n}\, {M}_{\!\scriptscriptstyle \chi^{\mbox{-}}_n}
 \mbox{\boldmath $U$}^{\!*}_{\!4n} \;
F_6\!\!\left({{M}_{\!\scriptscriptstyle \chi^{\mbox{-}}_{n}}^2
\over M_{\!\scriptscriptstyle S_{m}}^2} \right)  \; {\cal
D}^{s}_{\!3m} \, {\cal D}^{s*}_{\!2m} \;
%\lambda_{kh\scriptscriptstyle 1}
\frac{y_{\!\scriptscriptstyle e_2}}{g_{\scriptscriptstyle 2}}\;,
\eeq where we used the relation $\mbox{\boldmath
$U$}^{\!*}_{\!(j'+2)3} \sim \,\delta_{j',1}$ and $\mbox{\boldmath
$V$}_{\!(j+2)4} \sim \, \delta_{j,2}$. As mentioned above, the
expected $B_1^*\, \mu_{\scriptscriptstyle 2}$ combination comes
up. It clearly has a muon Yukawa suppression. Likewise, the
expression (\ref{2murl2g2}) becomes \beq \label{2murl2}
\sum_{n=1}^5 \sum_{m=1}^{5}  \mbox{\boldmath $V$}^{\!*}_{\!\!1n}\,
{M}_{\!\scriptscriptstyle \chi^{\mbox{-}}_n}
 \mbox{\boldmath $U$}_{\!3n} \;
F_6\!\!\left({{M}_{\!\scriptscriptstyle \chi^{\mbox{-}}_{n}}^2
\over M_{\!\scriptscriptstyle S_{m}}^2} \right)  \; {\cal
D}^{s}_{\!2m} \, {\cal D}^{s*}_{\!4m} \;
%\lambda_{kh\scriptscriptstyle 1}
\frac{y_{\!\scriptscriptstyle e_1}}{g_{\scriptscriptstyle 2}}\; .
\eeq
 It clearly has the $\mu_1^*\, B_{\scriptscriptstyle 2}$ combination with
an electron Yukawa suppression, and is thus smaller than expression
(\ref{2murl1}). It is confirmed by our numerical calculation. The
result we mentioned above is also discussed in
Ref.~\cite{Cheung:2001sb}. The expression (\ref{2murl2g3}) becomes
\beq \label{2murl3} \sum_{n=1}^5 \sum_{m=1}^{5} - \mbox{\boldmath
$U$}^{\!*}_{\!23}\,\mbox{\boldmath $V$}_{\!\!1n}\,
{M}_{\!\scriptscriptstyle \chi^{\mbox{-}}_n}
 \mbox{\boldmath $U$}^{\!*}_{\!2n} \;
F_6\!\!\left({{M}_{\!\scriptscriptstyle \chi^{\mbox{-}}_{n}}^2
\over M_{\!\scriptscriptstyle S_{m}}^2} \right)  \; {\cal
D}^{s}_{\!2m} \, {\cal D}^{s*}_{\!4m} \;
\frac{y_{\!\scriptscriptstyle e_2}}{g_{\scriptscriptstyle 2}}\; ,
\eeq and we have the $\mu_1^*\, B_{\scriptscriptstyle 2}$ combination
with a muon Yukawa suppression.
%%%%%%%%%%%%%%%%%%%%%%%%%%%%%%%%%%%%%%
The expression (\ref{2murl2g4}) becomes \beq \label{2murl4}
\sum_{n=1}^5 \sum_{m=1}^{5} - \mbox{\boldmath
$U$}_{\!24}\,\mbox{\boldmath $V$}^{\!*}_{\!\!1n}\,
{M}_{\!\scriptscriptstyle \chi^{\mbox{-}}_n}
 \mbox{\boldmath $U$}_{\!2n} \;
F_6\!\!\left({{M}_{\!\scriptscriptstyle \chi^{\mbox{-}}_{n}}^2
\over M_{\!\scriptscriptstyle S_{m}}^2} \right)  \; {\cal
D}^{s}_{\!3m} \, {\cal D}^{s*}_{\!2m} \;
\frac{y_{\!\scriptscriptstyle e_1}}{g_{\scriptscriptstyle 2}}\; ,
\eeq and we have the $B_1^*\, \mu_{\scriptscriptstyle 2}$ combination
with an electron Yukawa suppression. We expect that this
contribution is smaller than in expression (\ref{2murl3}),
because it has a larger Yukawa suppression.
%%%%%%%%%%%%%%%%%%%%%%%%%%%%%%%%%%%%%%%%% in sum

In summary, using an approximate formula of mass eigenstate
couplings, one can only obtain the expression (\ref{2murl1}) and
(\ref{2murl2}). However, in this paper, we take all contributions
into account including diagrams with higssinos and winos on the
external legs. Therefore, additional contributions, such as the
expression (\ref{2murl3}) and (\ref{2murl4}) in this case, would
be obtained. If we consider the $B_1^*\, \mu_{\scriptscriptstyle
2}$ combination, the dominant term still comes from expression
(\ref{2murl1}), since the new contribution, expression
(\ref{2murl4}), has an electron Yukawa suppression, while
expression (\ref{2murl1}) only has a muon Yukawa suppression.
However, the situation would be totally changed in the combination
of $\mu_1^*\, B_{\scriptscriptstyle 2}$, since the expression
(\ref{2murl2}) has an electron Yukawa suppression while the new
contribution, expression (\ref{2murl3}), only has a muon Yukawa
suppression. As a result, our exact formula made the allowed
region of the $(B_{2},\mu_{\scriptscriptstyle 1})$ parameter space
more stringent than that in Ref.~\cite{Cheung:2001sb}. We give
contours of $Br(\mu\to e \,\gamma)$ in the real
$(B_{2},\mu_{\scriptscriptstyle 1})$ plane in  Fig.~\ref{MB2}. The
solid lines represent the results obtained by using the exact mass
eigenstate couplings, while the dashed lines stand for the results
obtained by using the approximate ones. Likewise, the contour of
$Br(\tau \to e\,\gamma)$ is given in the real
$(B_{3},\mu_{\scriptscriptstyle 1})$ plane in Fig.~\ref{MB4}. The
present experimental limit is also shown and the allowed region at
90\% C.L. is shaded.

The 90\% C.L. upper limit on $|{\mu^*}\,{B}|$ or  $|{\mu}\,{B^*}|$
type combinations (normalized by $|\mu_{\scriptscriptstyle 0}|^3$,
$|\mu_{\scriptscriptstyle 0}|=135\,\mbox{GeV}$) is given by
\[
\frac{|B_1^* \, \mu_{\scriptscriptstyle
2}|}{|\mu_{\scriptscriptstyle 0}|^3}
 < 6.5\times 10^{-7} \;, \;\;\;
\frac{|B_2 \, \mu^*_{\scriptscriptstyle
1}|}{|\mu_{\scriptscriptstyle 0}|^3}
 < 7.1\times 10^{-7} \;, \;\;\;
\]
\[
\frac{|B_1^* \, \mu_{\scriptscriptstyle
3}|}{|\mu_{\scriptscriptstyle 0}|^3}
  < 1.4\times 10^{-4} \;, \;\;\;
\frac{|B_3 \, \mu^*_{\scriptscriptstyle
1}|}{|\mu_{\scriptscriptstyle 0}|^3}
  < 1.5\times 10^{-4} \;, \;\;\;
\]
\begin{equation}
\label{upexact} \frac{|B_2^* \, \mu_{\scriptscriptstyle
3}|}{|\mu_{\scriptscriptstyle 0}|^3}
  < 1.1\times 10^{-4} \;, \;\;\;
\frac{|B_3 \, \mu^*_{\scriptscriptstyle
2}|}{|\mu_{\scriptscriptstyle 0}|^3}
  < 1.2\times 10^{-4} \;.
\end{equation}
For a direct contrast, we also give the {\em incorrect} upper
bounds obtained by using the approximate formula:
\[
\frac{|B_2 \, \mu^*_{\scriptscriptstyle
1}|}{|\mu_{\scriptscriptstyle 0}|^3}
 < 1.4\times 10^{-4} \;, \;\;\;
\frac{|B_3 \, \mu^*_{\scriptscriptstyle
1}|}{|\mu_{\scriptscriptstyle 0}|^3}
  < 3.5\times 10^{0} \;, \;\;\;
\frac{|B_3 \, \mu^*_{\scriptscriptstyle
2}|}{|\mu_{\scriptscriptstyle 0}|^3}
  < 2.3\times 10^{-3} \;.
\]
%%%%%%%%%%%%%%%%%%%%%%%%%%%%%%%%%%%%%%
We can see clearly that the ratio between the limit of
$\frac{|B_2\, \mu^*_{\scriptscriptstyle
1}|}{|\mu_{\scriptscriptstyle 0}|^3}$  and the correct one in
Eq.(\ref{upexact}) is approximately the ratio of
$\frac{y_{\!\scriptscriptstyle e_2}}{y_{\!\scriptscriptstyle
e_1}}$. Therefore, the difference of the experimental bounds
between the exact formula (solid line) and the approximate formula
(dashed line) in Fig.~\ref{MB2} is the result of different Yukawa
suppressions, which is consistent with our analysis.
If we take the expected improvement from the MEG experiment into
account and assume $Br(\mu\to e \,\gamma) < 10^{-14} $,  we get
\[
\frac{|B_1^* \, \mu_{\scriptscriptstyle
2}|}{|\mu_{\scriptscriptstyle 0}|^3}
 < 1.2\times 10^{-8} \;, \;\;\;
\frac{|B_2 \, \mu^*_{\scriptscriptstyle
1}|}{|\mu_{\scriptscriptstyle 0}|^3}
 < 1.3\times 10^{-8} \;. \;\;\;
\]
Notice that this constraint is even more stringent than the one
from neutrino masses as you can see in Fig.~\ref{MB2}.

%=========================================================================
\subsection{Parameter variations}
%=========================================================================
In this section, we illustrate the effects of varying the input
SUSY parameters on the bounds, using $|\mu_{\scriptscriptstyle
1}^*\,\lambda_{\scriptscriptstyle 121}|$ and
$|\mu_{\scriptscriptstyle 1}^*\,B_{\scriptscriptstyle 2}|$ as
examples. The results are summarized in Table~\ref{table3}.
%1.
In the table, we list the variation of the
$\mu_{\scriptscriptstyle 0}$ and the $M_{\scriptscriptstyle 1}$ in
parts i and ii, respectively. Our numerical results show that the
bound is most stringent for small $|\mu_{\scriptscriptstyle 0}|$
in both $|\mu_{\scriptscriptstyle
1}^*\,\lambda_{\scriptscriptstyle 121}|$ and
$|\mu_{\scriptscriptstyle 1}^*\,B_{\scriptscriptstyle 2}|$ cases.
In addition, the increase of $M_{\scriptscriptstyle 1}$ also
weakens the bound. These results are reasonable since increasing
$\mu_{\scriptscriptstyle 0}$ and $M_{\scriptscriptstyle 1}={1\over
2}\,M_{\scriptscriptstyle 2}$ essentially increases the chargino
and neutralino masses.
%2.
In the case of $|\mu_{\scriptscriptstyle
1}^*\,\lambda_{\scriptscriptstyle 121}|$, the dominant diagram
involves mainly the $\tilde{l}_{\scriptscriptstyle
2}^{\scriptscriptstyle 0}$, while the $|\mu_{\scriptscriptstyle
1}^*\, B_{\scriptscriptstyle 2}|$ case involves the mixing between
$\tilde{l}_{\scriptscriptstyle 2}^{\scriptscriptstyle 0}$ and
$\tilde{l}_{\scriptscriptstyle 0}^{\scriptscriptstyle 0}$.
Therefore, varying $\tilde{m}^2_{\!\scriptscriptstyle E}$ does not
have much effect on the bounds while varying the corresponding
entries in $\tilde{m}^2_{\!\scriptscriptstyle L}$ changes the
bounds significantly (see  parts iii and iv).

%3.
Finally, part v of Table~\ref{table3} shows the $\tan\!\beta$
dependence of the results. From the table we can see that varying
$\tan\!\beta$ has only a little effect on
$|\mu_{\scriptscriptstyle 1}^*\,\lambda_{\scriptscriptstyle 121}|$,
but a rather significant effect on $|\mu_{\scriptscriptstyle
1}^*\, B_{\scriptscriptstyle 2}|$.
%%%%%%%%%%%%%%%%%%%%%%
Although the subdominant contributions mentioned in
Sec. IV B involve the Yukawa couplings, and thus have the
$\frac{1}{\cos \zb}$ dependence, the dominant contributions do not
have the $\tan\!\beta$ dependence \cite{Cheung:2001sb}. As a
result, the lack of sensitivity to $\tan\!\beta$ in the former
case is to be expected.
%%%%%%%%%%%%%%%%%%%%%%%%%
In the latter case, the numerical result shows that the bound has
a strong dependence on $\tan\!\beta$. There are two sources that
result in this dependence. The first one is the ${1\over
\cos\!\beta}$ dependence of the Yukawa coupling. The other is the
explicit $\tan\,\beta$ dependence of the dominant terms mentioned
in Sec. IV C. Figure~\ref{varytanb} shows a contour plot
of the experimental bound of $B(\mu \to e \,\gamma)$ in the (real)
plane of (${B_{\scriptscriptstyle 2}},{\mu_{\scriptscriptstyle
1}}$) for various values of $\tan\!\beta$. It not only shows
the $\tan\!\beta$ dependence of the results, but also illustrates
that the experimental bound gives a more stringent constraint in the
large $\tan\!\beta$ region.

%%%%%%%%%%%%%%%%%%%%%%%%%%%%%%%%%%%%%%%%%%%%%%%%%%%%%%%%%%%%%%%%%%%%%%
\section{Conclusion}
%%%%%%%%%%%%%%%%%%%%%%%%%%%%%%%%%%%%%%%%%%%%%%%%%%%%%%%%%%%%%%%%%%%%%%
In this paper, we have given explicit formulas and detailed
discussions on the full one-loop contribution to the radiative
decay of $\mu$ and $\tau$, namely, $\mu \to e\, \gamma$, $\tau \to
e\,\gamma$, and $\tau \to\mu \gamma$ for the generic supersymmetric
SM (without R parity). We use the exact formula of the mass
eigenstate couplings to calculate the branching ratios of these
leptonic radiative decay processes and compare them with the
results obtained in an earlier approximation  in Ref.
\cite{Cheung:2001sb}.
% 1
In some combinations of RPV parameters, the results obtained by
using these two approaches are exactly the same such as $\mu_1
\,\lambda_{\scriptscriptstyle 123}^*$, $\mu_1^*
\,\lambda_{\scriptscriptstyle 132}$, $\mu_2
\,\lambda_{\scriptscriptstyle 213}^*$, $\mu_2^*
\,\lambda_{\scriptscriptstyle 231}$, $\mu_3^*
\,\lambda_{\scriptscriptstyle 321}$, $\mu_3
\,\lambda_{\scriptscriptstyle 312}^*$, and all ${B^*\, \lambda}$
and ${B \, \lambda^*}$ type combinations.
% 2
The dominant terms of the other $\mu^*\lambda$ or $\mu \lambda^*$
combinations such as $\mu_2 \,\lambda_{\scriptscriptstyle 212}^*$,
$\mu_1^* \,\lambda_{\scriptscriptstyle 121}$, $\mu_3
\,\lambda_{\scriptscriptstyle 313}^*$, $\mu_1^*
\,\lambda_{\scriptscriptstyle 131}$, $\mu_3^*
\,\lambda_{\scriptscriptstyle 323}$, and $\mu_2
\,\lambda_{\scriptscriptstyle 232}^*$ are still the same in the
two methods, but the subdominant terms come from a new contribution.
% 3
In the ${B^*\, \mu}$ or ${B \, \mu^*}$ contributions, the dominant
contributions of $B_{\scriptscriptstyle 1}^* \,
\mu_{\scriptscriptstyle 2}$, $B_{\scriptscriptstyle 1}^*\,
\mu_{\scriptscriptstyle 3}$, and $B_{\scriptscriptstyle 2}^*\,
\mu_{\scriptscriptstyle 3}$ are the same in the exact and
approximate formulas. However, the dominant contributions of
$\mu_{\scriptscriptstyle 1}^* \, B_{\scriptscriptstyle 2}$,
$\mu_{\scriptscriptstyle 1}^* \, B_{\scriptscriptstyle 3}$, and
$\mu_{\scriptscriptstyle 2}^* \, B_{\scriptscriptstyle 3}$ are
totally new. The upper bound on these combinations obtained from
the experimental limit are
\[
\frac{|B_2 \, \mu^*_{\scriptscriptstyle
1}|}{|\mu_{\scriptscriptstyle 0}|^3}
 < 7.1\times 10^{-7} \;, \;\;\;
\]
\[
\frac{|B_3 \, \mu^*_{\scriptscriptstyle
1}|}{|\mu_{\scriptscriptstyle 0}|^3}
  < 1.5\times 10^{-4} \;, \;\;\;
\]
\[
\frac{|B_3 \, \mu^*_{\scriptscriptstyle
2}|}{|\mu_{\scriptscriptstyle 0}|^3}
  < 1.2\times 10^{-4} \;. \;\;\;
\]
As a result, our exact formulas impose more stringent constraints
on the admissible region of parameter spaces for the GSSM, or SUSY
without R parity.
If we also consider the expected improvement from the MEG
experiment and assume $Br(\mu\to e \,\gamma) < 10^{-14} $, we
could get even more stringent constraints on the
$B_{\scriptscriptstyle 2}^*\, \mu_{\scriptscriptstyle 1}$
combination
\[
\frac{|B_2 \, \mu^*_{\scriptscriptstyle
1}|}{|\mu_{\scriptscriptstyle 0}|^3}
 < 1.3\times 10^{-8} \;. \;\;\;
\]
The constraint is even more stringent than the naive constraint
from neutrino masses imposed here, which indicates a very encouraging
scenario for future probing of the leptonic radiative decay and may
be as well as the $\tau$ decays. A more involved analysis will have to
be performed on the full model parameter space matching the
radiative decays to neutrino mass generations. We want to
highlight though that the scale for the actual neutrino masses is
not expected to be reduced while probes on the leptonic radiative
decays, and for that matter the other lepton number/flavor
violating decays, can be improved. That makes the latter a
promising ground to further explore models like SUSY without R
parity with rich lepton number/flavor violating structures.

%%%%%%%%%%%%%%%%%%%%%%%%%%%%%%%%%
\acknowledgments The work of O. K. is partially supported by
research Grant No.96-2112-M-008-007-MY3 of the NSC of Taiwan.
The authors would like to thank Katherine Sutton for
 editing this paper.
%%%%%%%%%%%%%%%%%%%%%%%%%%%%%%%%%%%%%%%%%%%%%%%%%%%%%%%%%%
% The end
%%%%%%%%%%%%%%%%%%%%%%%%%%%%%%%%%%%%%%%%%%%%%%%%%%%%%%%%%%

%%%%%%%%%%%%%%%%%%%%%%%%%%%%%%%%%%%%%%%%%%%%%%%%%%%%%%%%%%

%%%%%%%%%%%%%%%%%%%%%%%%%%%%%%%%%%%%%%%%%%%%%%%%
%\newpage
%%%%%%%%%%%%%%%%%%%%% Figures %%%%%%%%%%%%%%%%%%%%%%
\begin{figure}
\begin{center}
%{\bf FIGURES}
\bigskip
%\vspace{1cm}
\includegraphics[scale=0.9]{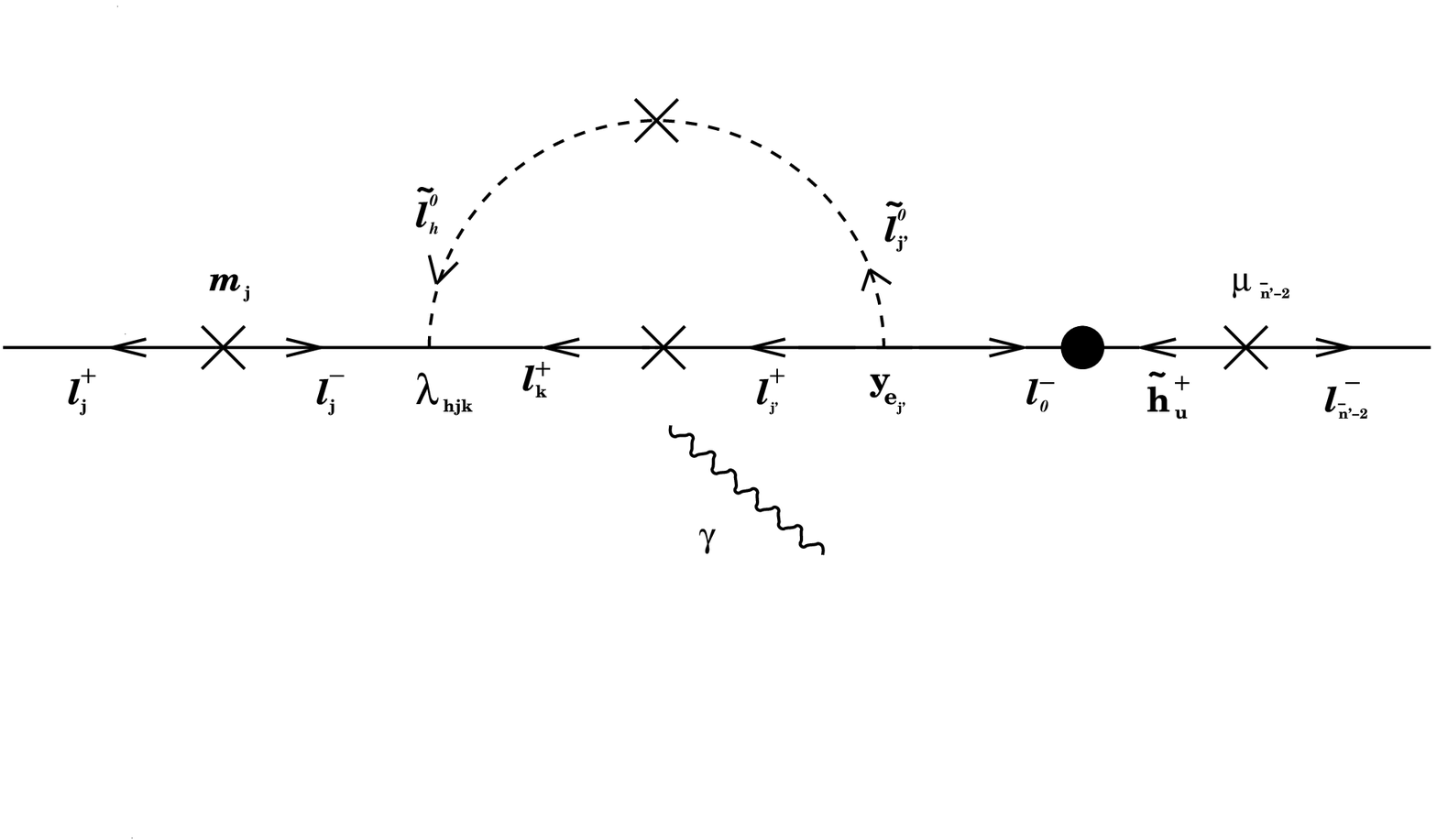}
\caption{Diagram with chirality flip on the external line, which
has larger Yukawa suppression.} \label{00Chen10}
\includegraphics[scale=1.0]{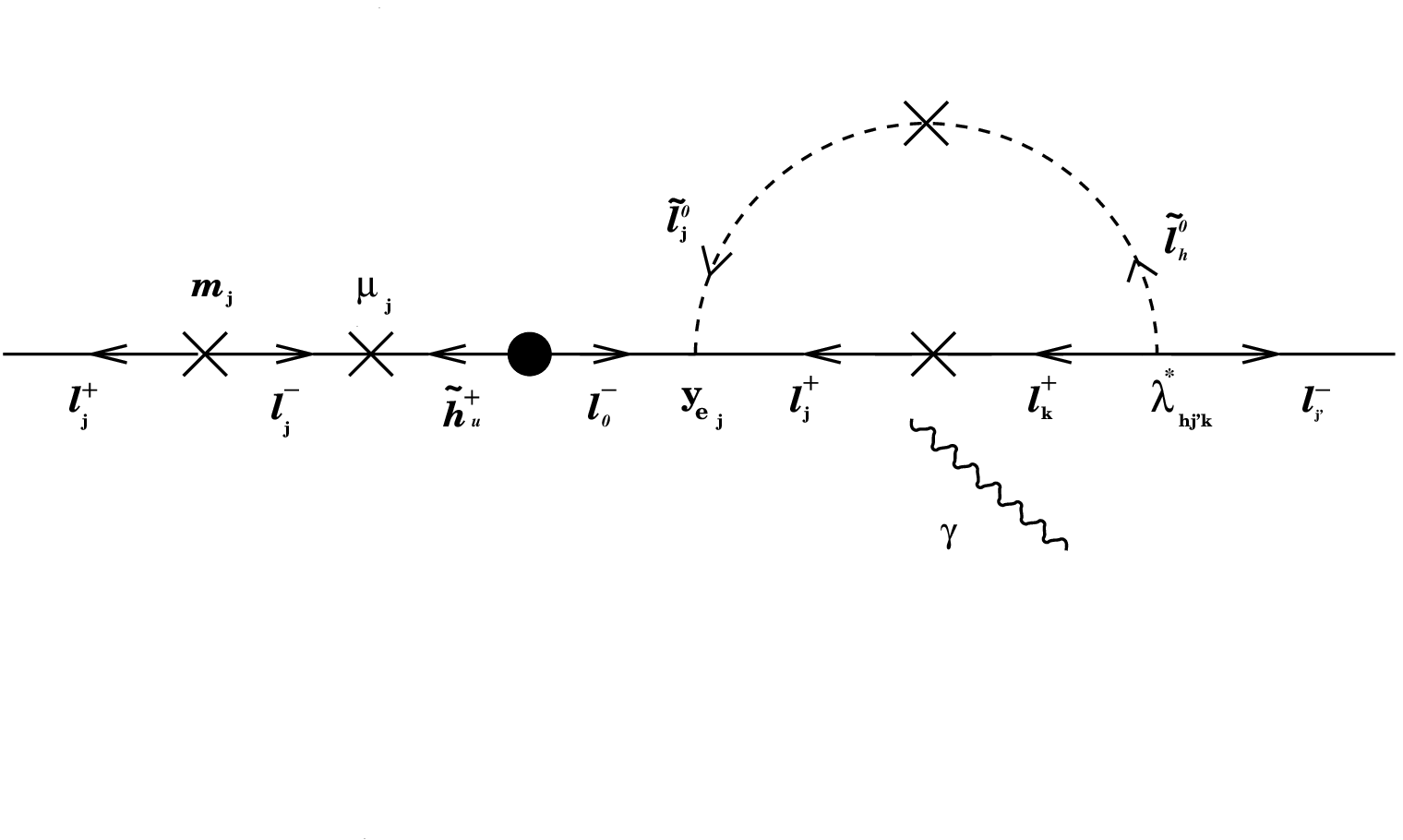}
\caption{Diagram with chirality flip on the external line, which
has smaller Yukawa suppression.} \label{00Chen20}
\end{center}
\end{figure}

%%%%%%%%%%%%%%%%%%%%%%%%%%%%%%%%%%%%%%
\begin{figure}
\begin{center}
\includegraphics[scale=1.0]{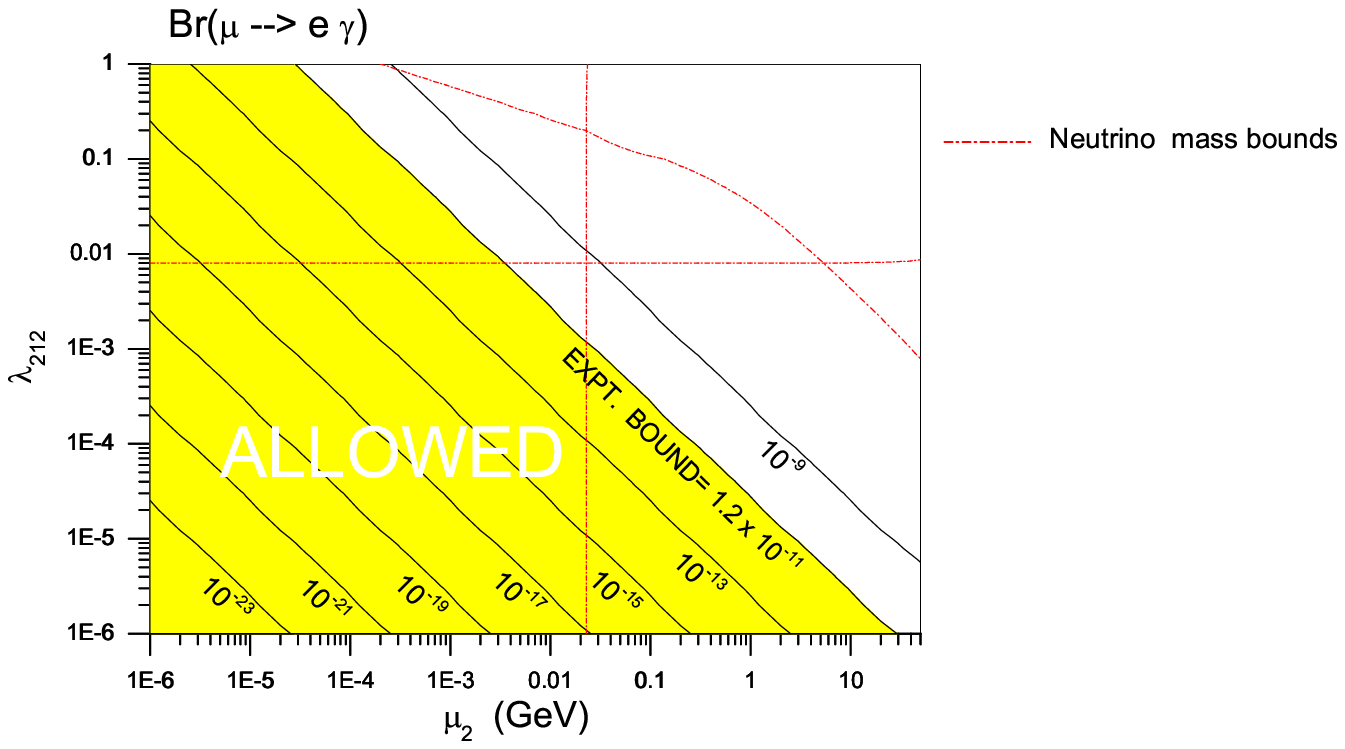}
\caption{Contours of $B(\mu \to e \,\gamma)$ in the (real) plane
of (${\mu_{\scriptscriptstyle 2}},{\lambda_{\scriptscriptstyle
212}} $). The 90\% C.L. allowed region is shaded.
The dash-dotted (red) lines are sub-eV neutrino mass bounds.}\label{ML1}
\end{center}
\end{figure}

%%%%%%%%%%%%%%%%%%%%%%%%%%%%%%%%%%%%%%
\begin{figure}
\begin{center}
% \vspace{1cm}
\includegraphics[scale=1.0]{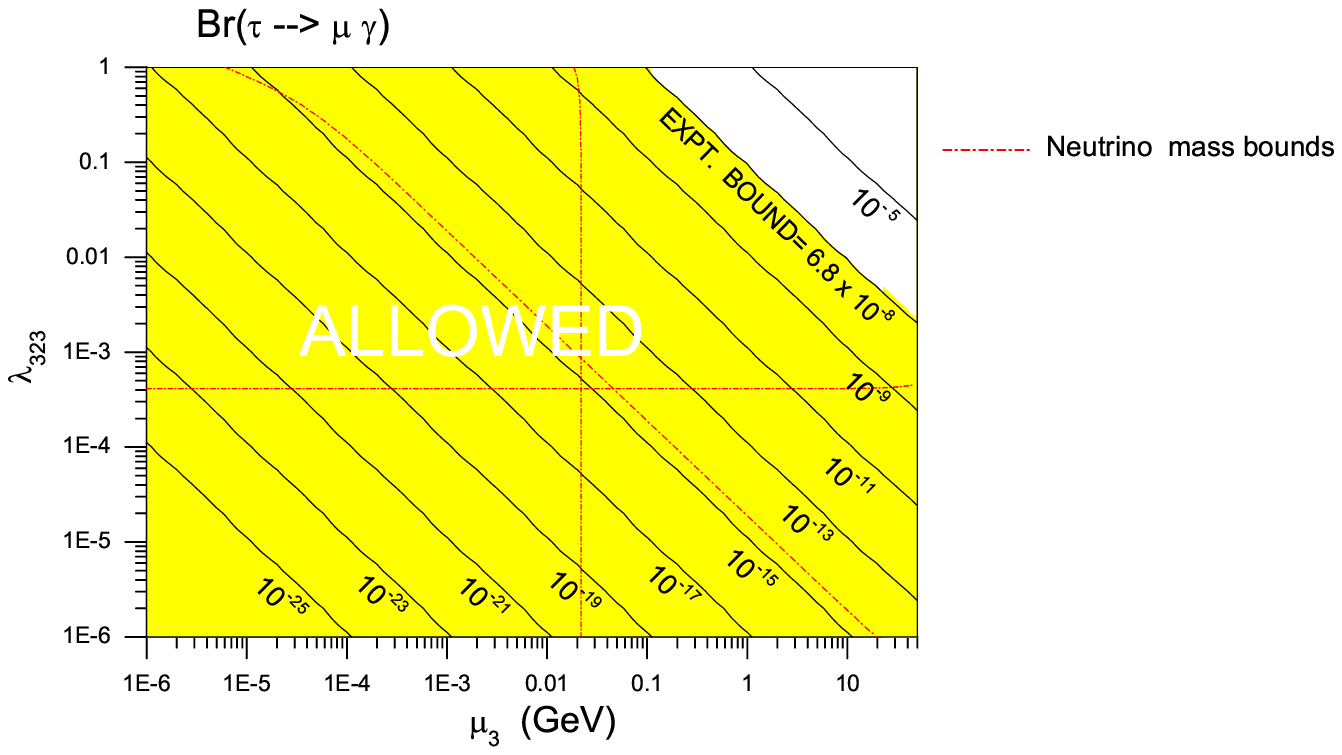}
\caption{Contours of $B(\tau \to \mu \,\gamma)$ in the (real)
plane of (${\mu_{\scriptscriptstyle
3}},{\lambda_{\scriptscriptstyle 323}} $). The 90\% C.L. allowed
region is shaded.
The dash-dotted (red) lines are sub-eV neutrino mass bounds.} \label{ML6}
\includegraphics[scale=1.0]{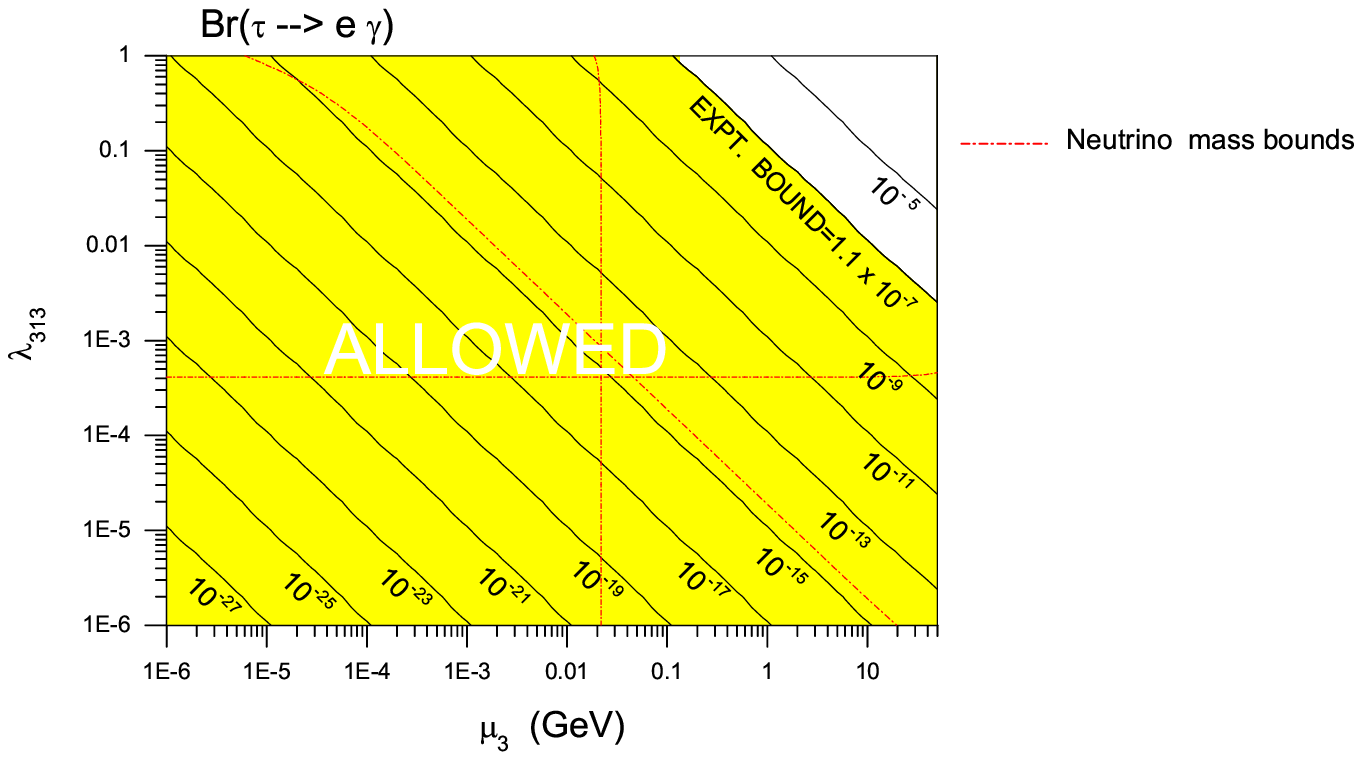}
\caption{Contours of $B(\tau \to e \,\gamma)$ in the (real) plane
of (${\mu_{\scriptscriptstyle 3}},{\lambda_{\scriptscriptstyle
313}} $). The 90\% C.L. allowed region is shaded.
The dash-dotted (red) lines are sub-eV neutrino mass bounds.} \label{ML10}
\end{center}
\end{figure}
%%%%%%%%%%%%%%%%%%%%%%%%%%%%%%%%%%%%%
\begin{figure}
\begin{center}
% \vspace{1cm}
\includegraphics[scale=0.9]{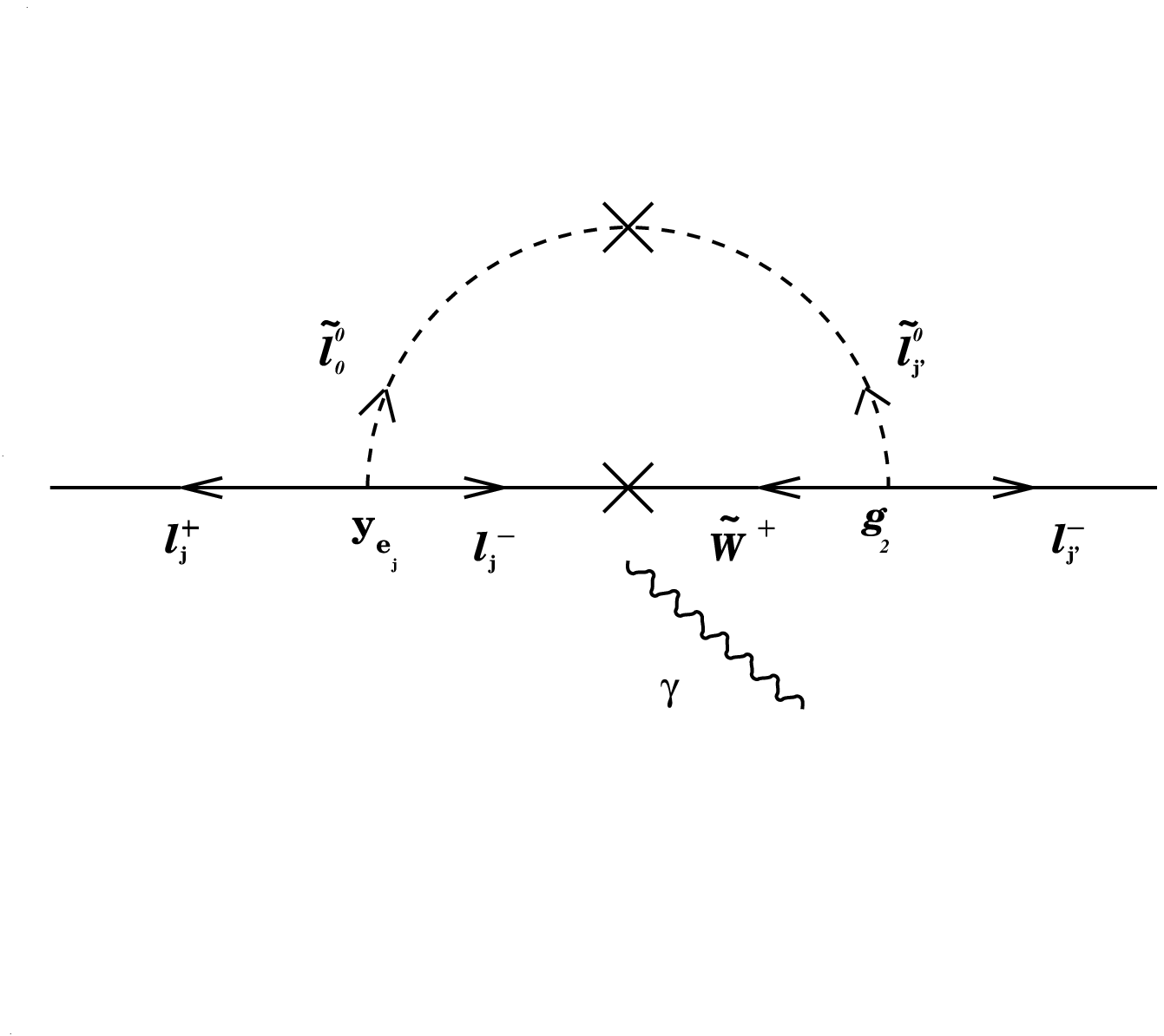}
\caption{The charginolike loop diagram contributes to the
leptonic radiative decay due to $B_{j'}^*\, \mu_j$
combination.}\label{00Chen30}
\includegraphics[scale=0.9]{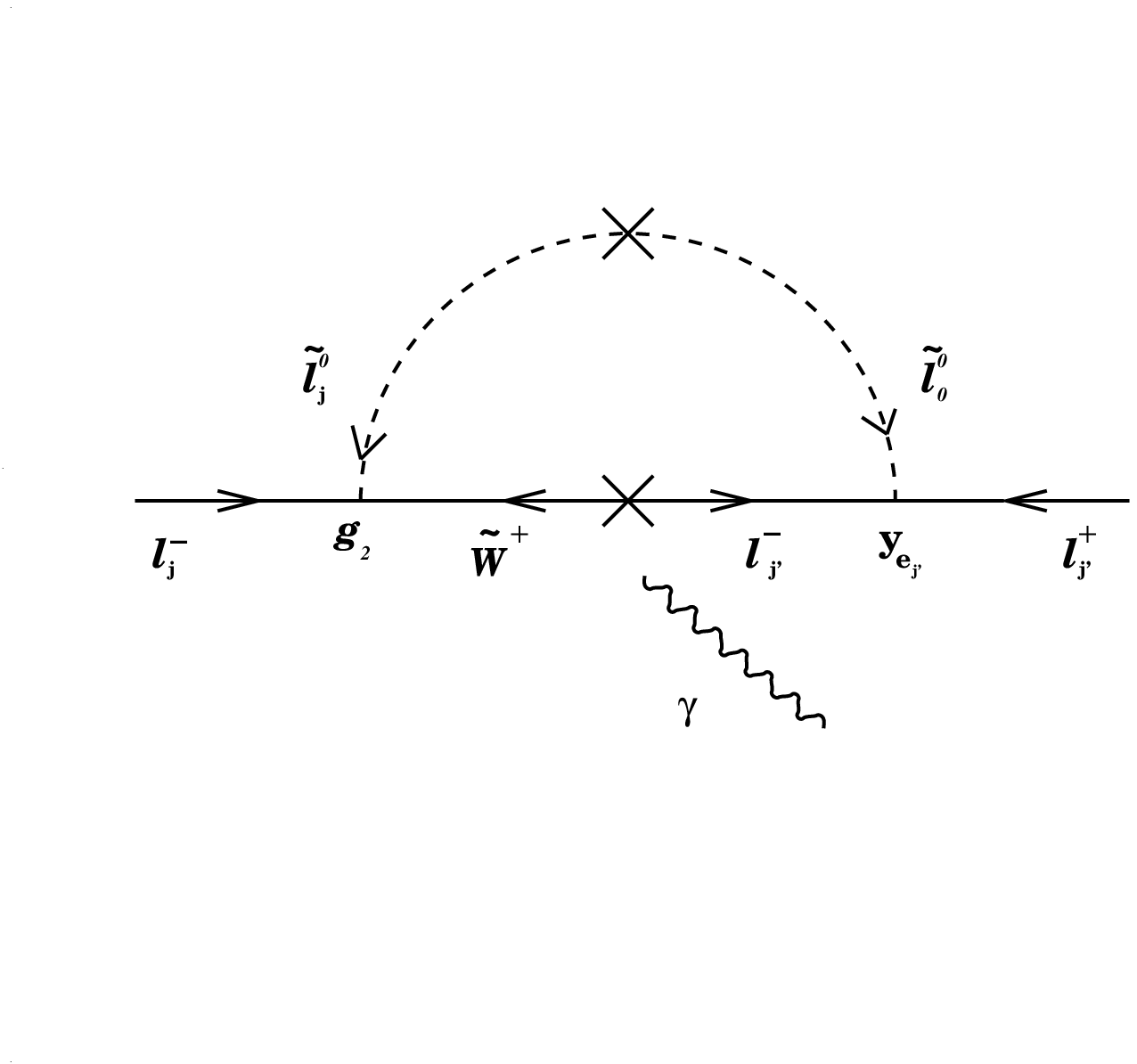}
\caption{The charginolike loop diagram contributes to the
leptonic radiative decay due to $\mu_{j'}^*\, B_j$
combination.}\label{00Chen42}
\end{center}
\end{figure}
%%%%%%%%%%%%%%%%%%%%%%%%%%%%%%%%%%%%%%%%%
\begin{figure}
\begin{center}
% \vspace{1cm}
\includegraphics[scale=0.9]{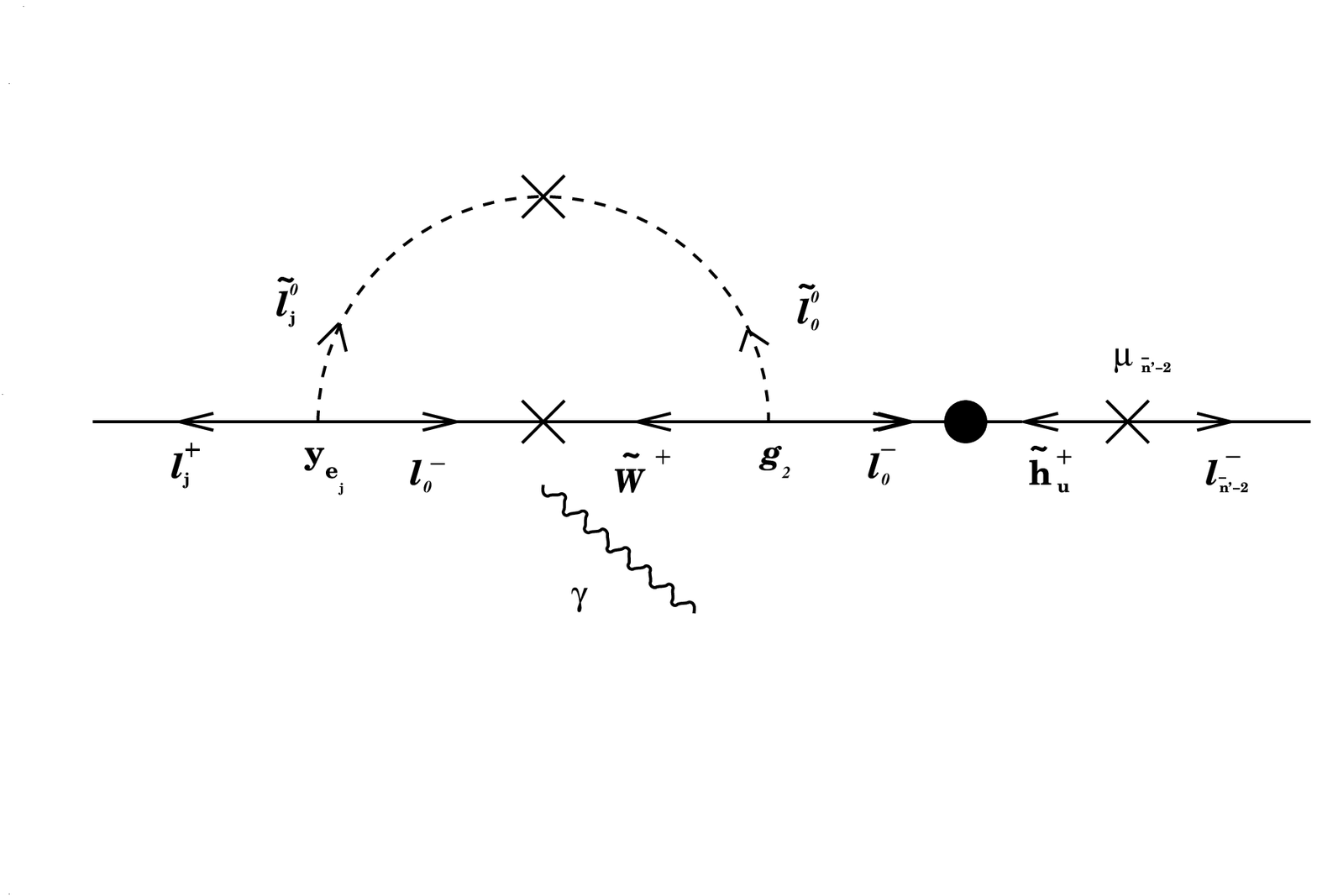}
\caption{The charginolike loop diagram contributes to the
leptonic radiative decay due to $B_j \,\mu_{\scriptscriptstyle
{\bar{n}'-2}}^*$ combination, where ${\bar{n}'-2}<j$}
\label{00Chen50}
\includegraphics[scale=0.9]{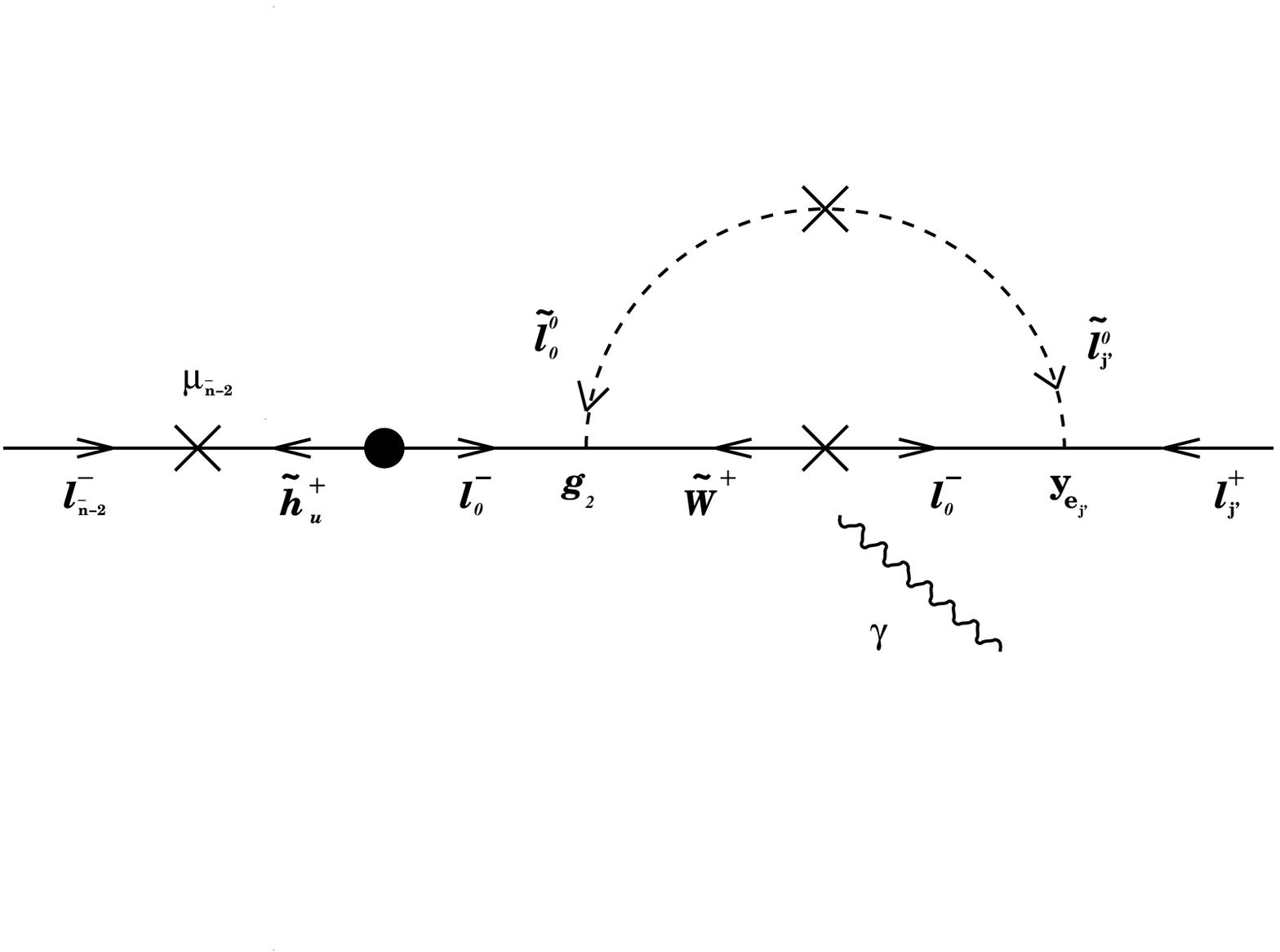}
\caption{The charginolike loop diagram contributes to the
leptonic radiative decay due to $B_{j'}^*
\,\mu_{\scriptscriptstyle {\bar{n}-2}}$ combination, where
$j'<{\bar{n}-2}$} \label{00Chen60}
\end{center}
\end{figure}
%%%%%%%%%%%%%%%%%%%%%%%%%%%%%%%%%%%%%%%%%
%%%%%%%%%%%%%%%%%%%%%%%%%%%%%%%%%%%%%%%%
\begin{figure}
\begin{center}
% \vspace{1cm}
\includegraphics[scale=1.0]{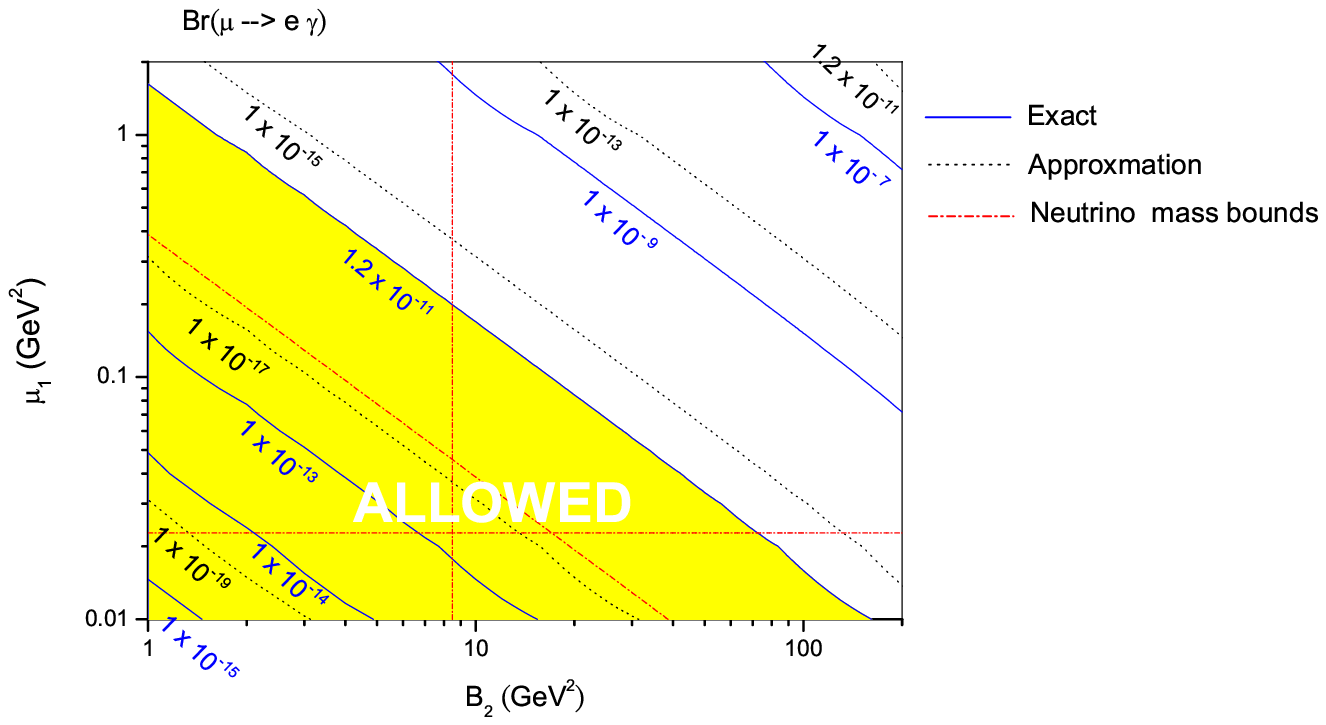}
\caption{Contours of $B(\mu \to e \,\gamma)$ in the (real) plane
of (${B_{\scriptscriptstyle 2}},{\mu_{\scriptscriptstyle 1}} $).
The 90\% C.L. allowed region is shaded. The solid lines represent
the results obtained by using the exact mass eigenstate couplings,
while the dashed lines stand for the results obtained by using the
approximate ones. The dash-dotted (red) lines are neutrino mass
bounds. Notice the MEG experiment targets probing the decay
at $10^{-13}-10^{-14}$.} \label{MB2}
\includegraphics[scale=1.0]{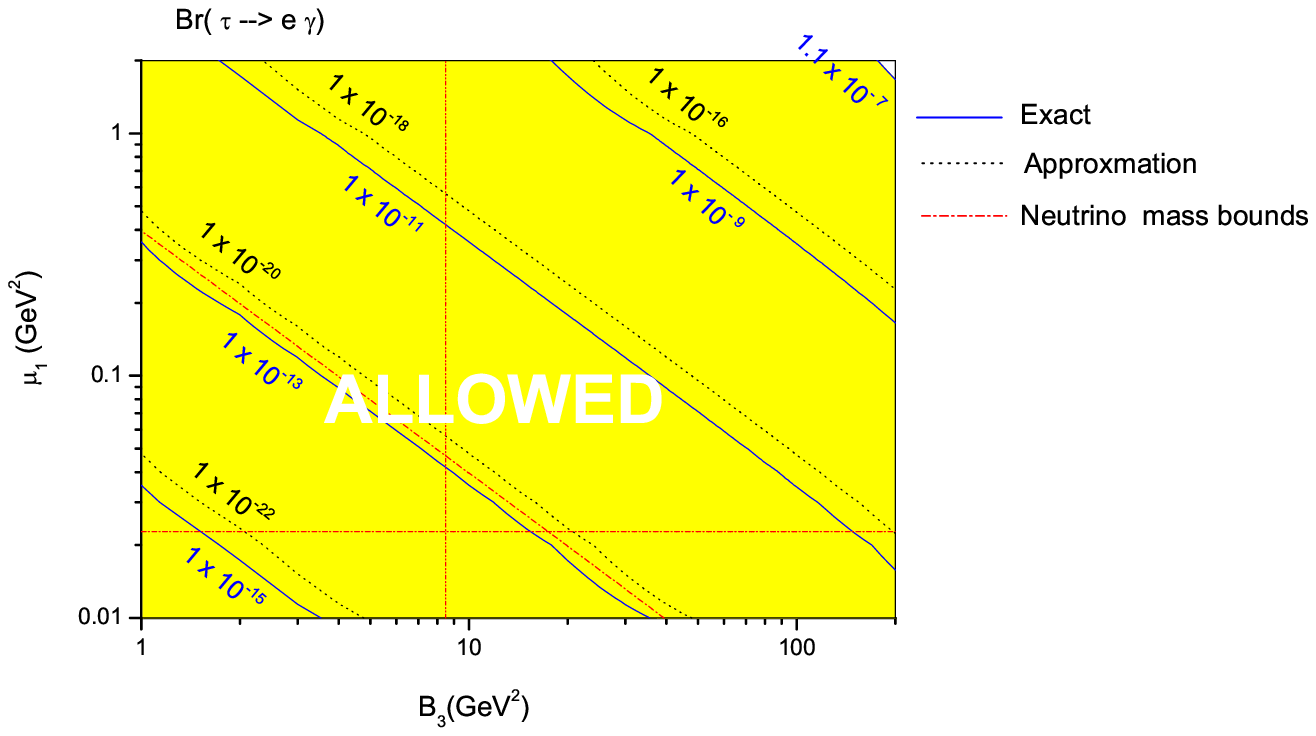}
\caption{Contours of $B(\tau \to e \,\gamma)$ in the (real) plane
of (${B_{\scriptscriptstyle 3}},{\mu_{\scriptscriptstyle 1}} $).
The 90\% C.L. allowed region is shaded. The solid lines represent
the results obtained by using the exact mass eigenstate couplings,
while the dashed lines stand for the results obtained by using the
approximate ones. The dash-dotted (red) lines are neutrino mass
bounds.} \label{MB4}
\end{center}
\end{figure}
%%%%%%%%%%%%%%%%%%%%%%%%%%%%%%%%%%%%%%%
\begin{figure}
\begin{center}
\includegraphics[scale=1.0]{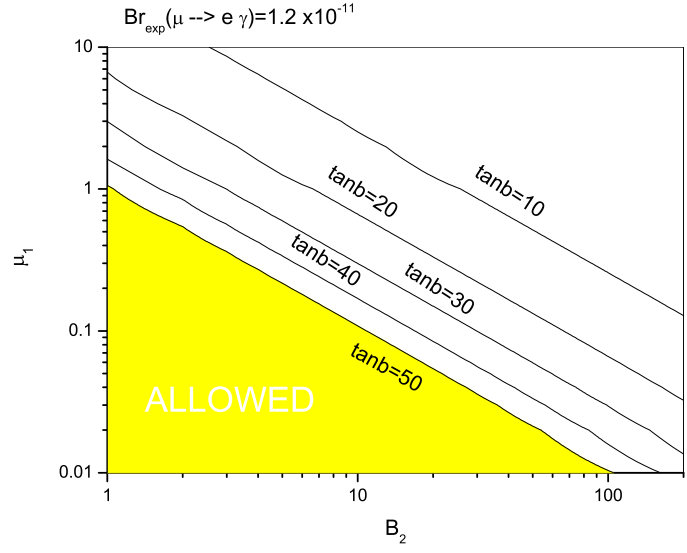}
\caption{Contours of the experimental bound of $B(\mu \to e
\,\gamma)$ in the (real) plane of (${B_{\scriptscriptstyle
2}},{\mu_{\scriptscriptstyle 1}}$) with different values of
$\tan\!\beta$, ranging from 10 to 50, among which $\tan\!\beta$=50
gives the most stringent constraint and the corresponding 90\%
C.L. allowed region is shaded. } \label{varytanb}
\end{center}
\end{figure}

%%%%%%%%%%%%%%%%%%%%%% Tables %%%%%%%%%%%%%%%%%%%%
%\newpage
%----------------------------------------------------------table 1
\begin{table}[bh]
\vspace*{.5in} \caption{\small \label{table1} Basic input SUSY
parameters for the numerical results presented. These values are
adopted unless otherwise specified.}
\bigskip
\centering
\begin{tabular}{cccc}
$M_{\scriptscriptstyle 1}$ (GeV) & $M_{\scriptscriptstyle 2}$ (GeV)  & $\mu_{\scriptscriptstyle 0}$ (GeV)  & $\tan\!\beta$ \\
\hline
100  & 200 & 135 & 40 \\
\hline \hline $\tilde{m}^2_{\!{\scriptscriptstyle L}}$ ($10^4$
GeV$^{2}$) & $\tilde{m}^2_{\!{\scriptscriptstyle E}}$ ($10^4$
GeV$^{2}$) &
$A_e$ (GeV) & \\
\hline
diag$\{2,1,1,1\}$ & diag$\{1,1,1\}$ & 100 & \\
\end{tabular}
\end{table}

%%%%%%%%%%%%%%%%%%%%%%%%%%%%%%%%%%%%%%%%%%
%----------------------The summary of  bounds-------------------------------------------
\begin{table}[bh]
\caption{\small \label{table2} Summary of bounds on various
combinations of two R-parity violating parameters, normalized by
$|\mu_{\scriptscriptstyle 0}|=135$ GeV. The input parameters are
as in Table~\ref{table1}.}
\bigskip
\centering
\begin{tabular}{ll}
%--------------------------------- mu lambda
\ \ $\frac{|{\mu_{\scriptscriptstyle
3}^*}\,{\lambda_{\scriptscriptstyle 321}}|}
{|\mu_{\scriptscriptstyle 0}|}\;, \;\;\;
\frac{|{\mu_{\scriptscriptstyle
1}^*}\,{\lambda_{\scriptscriptstyle 121}}|}
{|\mu_{\scriptscriptstyle 0}|}\;, \;\;\;
\frac{|{\mu_{\scriptscriptstyle 3}}\,{\lambda_{\scriptscriptstyle
312}^*}|} {|\mu_{\scriptscriptstyle 0}|}\;, \;\;\; \mbox{or}
\;\;\; \frac{|{\mu_{\scriptscriptstyle
2}}\,{\lambda_{\scriptscriptstyle 212}^*}|}
{|\mu_{\scriptscriptstyle 0}|}\; \;\;\;$ & $< 2.1 \times 10^{-7}$
\ \
\\ \hline
\ \ $\frac{|{\mu_{\scriptscriptstyle
2}^*}\,{\lambda_{\scriptscriptstyle 232}}|}
{|\mu_{\scriptscriptstyle 0}|}\;, \;\;\;
\frac{|{\mu_{\scriptscriptstyle
1}^*}\,{\lambda_{\scriptscriptstyle 132}}|}
{|\mu_{\scriptscriptstyle 0}|}\;, \;\;\;
\frac{|{\mu_{\scriptscriptstyle 3}}\,{\lambda_{\scriptscriptstyle
323}^*}|} {|\mu_{\scriptscriptstyle 0}|}\;, \;\;\; \mbox{or}
\;\;\; \frac{|{\mu_{\scriptscriptstyle
1}}\,{\lambda_{\scriptscriptstyle 123}^*}|}
{|\mu_{\scriptscriptstyle 0}|}\; \;\;\;$ & $< 7.0 \times 10^{-4}$
\ \
\\ \hline
\ \ $\frac{|{\mu_{\scriptscriptstyle
2}^*}\,{\lambda_{\scriptscriptstyle 231}}|}
{|\mu_{\scriptscriptstyle 0}|}\;, \;\;\;
\frac{|{\mu_{\scriptscriptstyle
1}^*}\,{\lambda_{\scriptscriptstyle 131}}|}
{|\mu_{\scriptscriptstyle 0}|}\;, \;\;\;
\frac{|{\mu_{\scriptscriptstyle 3}}\,{\lambda_{\scriptscriptstyle
313}^*}|} {|\mu_{\scriptscriptstyle 0}|}\;, \;\;\; \mbox{or}
\;\;\; \frac{|{\mu_{\scriptscriptstyle
2}}\,{\lambda_{\scriptscriptstyle 213}^*}|}
{|\mu_{\scriptscriptstyle 0}|}\; \;\;\;$ & $< 8.5 \times 10^{-4}$
\ \
\\ \hline
\ \
%--------------------------------- b mu
$\frac{|B_1^* \, \mu_{\scriptscriptstyle
2}|}{|\mu_{\scriptscriptstyle 0}|^3}$ & $< 6.5\times 10^{-7}$ \ \
\\ \hline
\ \ $\frac{|B_2 \, \mu^*_{\scriptscriptstyle
1}|}{|\mu_{\scriptscriptstyle 0}|^3}$
 & $< 7.1\times 10^{-7}$ \ \
\\ \hline
\ \
%%%%%%
$\frac{|B_1^* \, \mu_{\scriptscriptstyle
3}|}{|\mu_{\scriptscriptstyle 0}|^3}$
 & $< 1.4\times 10^{-4}$ \ \
\\ \hline
\ \ $\frac{|B_3 \, \mu^*_{\scriptscriptstyle
1}|}{|\mu_{\scriptscriptstyle 0}|^3}$
 & $< 1.5\times 10^{-4}$ \ \
\\ \hline
\ \
%%%%%
$\frac{|B_2^* \, \mu_{\scriptscriptstyle
3}|}{|\mu_{\scriptscriptstyle 0}|^3}$
 & $< 1.1\times 10^{-4}$ \ \
\\ \hline
\ \ $\frac{|B_3 \, \mu^*_{\scriptscriptstyle
2}|}{|\mu_{\scriptscriptstyle 0}|^3}$
 & $< 1.2\times 10^{-4}$ \ \
\\ \hline
\ \
%--------------------------------- mu mu
$\frac{|\mu_{\scriptscriptstyle 1}^*\, \mu_{\scriptscriptstyle
2}|} {|\mu_{\scriptscriptstyle 0}|^2}$   &   $ < 3.7 \times 10
^{-5}$ \ \
\\ \hline
\ \ $\frac{|\mu_{\scriptscriptstyle 1}^*\, \mu_{\scriptscriptstyle
3}|} {|\mu_{\scriptscriptstyle 0}|^2}$   &   $ < 4.7 \times 10
^{-3}$ \ \
\\ \hline
\ \ $\frac{|\mu_{\scriptscriptstyle 2}^*\, \mu_{\scriptscriptstyle
3}|} {|\mu_{\scriptscriptstyle 0}|^2}$   &   $ < 3.6 \times 10
^{-3}$ \ \
\\ \hline
\ \
%--------------------------------- b lambda
$\frac{|{B_{\scriptscriptstyle 3}^*}\,{\lambda_{\scriptscriptstyle
321}}|} {|\mu_{\scriptscriptstyle 0}|^2}\;, \;\;\;
\frac{|{B_{\scriptscriptstyle 1}^*}\,{\lambda_{\scriptscriptstyle
121}}|} {|\mu_{\scriptscriptstyle 0}|^2}\;, \;\;\;
\frac{|{B_{\scriptscriptstyle 3}}\,{\lambda_{\scriptscriptstyle
312}^*}|} {|\mu_{\scriptscriptstyle 0}|^2}\;, \;\;\; \mbox{or}
\;\;\; \frac{|{B_{\scriptscriptstyle
2}}\,{\lambda_{\scriptscriptstyle 212}^*}|}
{|\mu_{\scriptscriptstyle 0}|^2}\; \;\;\;$ & $< 1.3 \times
10^{-4}$ \ \
\\ \hline
\ \ $\frac{|{B_{\scriptscriptstyle
2}^*}\,{\lambda_{\scriptscriptstyle 232}}|}
{|\mu_{\scriptscriptstyle 0}|^2}\;, \;\;\;
\frac{|{B_{\scriptscriptstyle 1}^*}\,{\lambda_{\scriptscriptstyle
132}}|} {|\mu_{\scriptscriptstyle 0}|^2}\;, \;\;\;
\frac{|{B_{\scriptscriptstyle 3}}\,{\lambda_{\scriptscriptstyle
323}^*}|} {|\mu_{\scriptscriptstyle 0}|^2}\;, \;\;\; \mbox{or}
\;\;\; \frac{|{B_{\scriptscriptstyle
1}}\,{\lambda_{\scriptscriptstyle 123}^*}|}
{|\mu_{\scriptscriptstyle 0}|^2}\; \;\;\;$ & $< 1.4 \times
10^{-3}$ \ \
\\ \hline
\ \ $\frac{|{B_{\scriptscriptstyle
2}^*}\,{\lambda_{\scriptscriptstyle 231}}|}
{|\mu_{\scriptscriptstyle 0}|^2}\;, \;\;\;
\frac{|{B_{\scriptscriptstyle 1}^*}\,{\lambda_{\scriptscriptstyle
131}}|} {|\mu_{\scriptscriptstyle 0}|^2}\;, \;\;\;
\frac{|{B_{\scriptscriptstyle 3}}\,{\lambda_{\scriptscriptstyle
313}^*}|} {|\mu_{\scriptscriptstyle 0}|^2}\;, \;\;\; \mbox{or}
\;\;\; \frac{|{B_{\scriptscriptstyle
2}}\,{\lambda_{\scriptscriptstyle 213}^*}|}
{|\mu_{\scriptscriptstyle 0}|^2}\; \;\;\;$ & $< 1.9 \times
10^{-3}$ \ \
\\ \hline
\ \
\end{tabular}
\end{table}

%-----------------------parameter variations-------------------------------------------------------------
%-----------------------------------table 3
\newpage
\begin{table}[bh]
\caption{\small \label{table3} Effects of parameter variations of
interest, on the bounds of ${|{\mu_{\scriptscriptstyle
1}^*}\,{\lambda_{\scriptscriptstyle 121}}|} \cdot
{(135\,\mbox{GeV})^{-1}}$ and ${|\mu_{\scriptscriptstyle 1}^*\,
B_{\scriptscriptstyle 2}|} \cdot {(135\,\mbox{GeV})^{-3}}$. Note
that the fixed mass scale of $135\,\mbox{GeV}$ is used for
normalization to extract numerical bounds. }
\bigskip
\centering
\begin{tabular}{llll}
{\large Parameter changes}  &  \multicolumn{2}{c}{\large Normalized numerical bounds} \\
 & \multicolumn{1}{c}{$\frac{|{\mu_{\scriptscriptstyle 1}^*}\,{\lambda_{\scriptscriptstyle 121}}|}{(135\; {\rm GeV})}$}
 & \multicolumn{1}{c}{$\frac{|\mu_{\scriptscriptstyle 1}^*\, B_{\scriptscriptstyle 2}|}{(135\; {\rm GeV})^3}$}\\
\hline Original inputs of Table \ref{table1} &
  $<2.1 \times 10^{-7}$  & $<7.1 \times 10^{-7}$   \\
%================================================================ I added this
\hline (i)\,\; $\tilde{m}^2_{\!{\scriptscriptstyle L}} = {\rm
diag}\{20000,500^2,500^2,500^2\}$ GeV$^{2}$
   &  &   \\
\mbox{\hspace{0.3in}}$\tilde{m}^2_{\!{\scriptscriptstyle E}} =
{\rm diag}\{500^2,500^2,500^2\}$ GeV$^{2}$
   &  &   \\
%=================================================================
\mbox{\hspace{0.3in}}$\mu_{\scriptscriptstyle 0}=500$ GeV
& $< 7.5\times 10^{-6}$ & $<2.1 \times 10^{-4}$  \\
\mbox{\hspace{0.3in}}$\mu_{\scriptscriptstyle 0}=250$ GeV
& $< 2.7\times 10^{-6}$ & $<3.0 \times 10^{-5}$  \\
\mbox{\hspace{0.3in}}$\mu_{\scriptscriptstyle 0}=135$ GeV
& $< 1.3\times 10^{-6}$ & $<7.1 \times 10^{-6}$  \\
\mbox{\hspace{0.3in}}$\mu_{\scriptscriptstyle 0}=-135$ GeV
& $< 1.3\times 10^{-6}$ & $<7.3\times 10^{-6}$  \\
\mbox{\hspace{0.3in}}$\mu_{\scriptscriptstyle 0}=-250$ GeV
& $< 2.9\times 10^{-6}$ & $<3.1 \times 10^{-5}$  \\
\mbox{\hspace{0.3in}}$\mu_{\scriptscriptstyle 0}=-500$ GeV
& $< 8.2\times 10^{-6}$ & $<2.1 \times 10^{-4}$  \\
\hline (ii) $M_{\scriptscriptstyle 1}={1\over
2}{M_{\scriptscriptstyle 2}}=500$ GeV
& $< 1.2\times 10^{-6}$ & $< 4.5\times 10^{-6}$  \\
\hline (iii) $\tilde{m}^2_{\!{\scriptscriptstyle L}}
=20000\times\; {\rm diag}\{1,1,1,1\}$ GeV$^{2}$
   & $< 2.9\times 10^{-7}$ & $< 9.7\times 10^{-7}$  \\
\mbox{\hspace{0.3in}}$\tilde{m}^2_{\!{\scriptscriptstyle L}} =
{\rm diag}\{20000,1000^2,1000^2,1000^2\}$ GeV$^{2}$
   & $< 3.0\times 10^{-6}$ & $< 2.1\times 10^{-5}$  \\
\hline (iv) $\tilde{m}^2_{\!{\scriptscriptstyle E}} =20000\times
\; {\rm diag}\{1,1,1\}$ GeV$^{2}$
    & $< 2.1\times 10^{-7}$ & $< 7.3\times 10^{-7}$ \\
\mbox{\hspace{0.3in}}$\tilde{m}^2_{\!{\scriptscriptstyle E}} =
{\rm diag}\{1000^2,1000^2,1000^2\}$ GeV$^{2}$
    & $< 2.2\times 10^{-7}$ & $< 8.7\times 10^{-7}$ \\
\hline (v)\, $\tilde{m}^2_{\!{\scriptscriptstyle L}} = {\rm
diag}\{20000,500^2,500^2,500^2\}$ GeV$^{2}$
   &  &   \\
\mbox{\hspace{0.3in}}$\tilde{m}^2_{\!{\scriptscriptstyle E}} =
{\rm diag}\{500^2,500^2,500^2\}$ GeV$^{2}$
   &  &   \\
%%%%%%%%%%%%%%%%%%%%%%%%%%%%%%%%%%%%%%
\mbox{\hspace{0.2in}} $\mu_{\scriptscriptstyle 0}=135$ GeV,
                      $\tan\!\beta$=2  & $< 8.7\times 10^{-7}$ & $<2.6 \times 10^{-3}$  \\
\mbox{\hspace{1.3in}} $\tan\!\beta$=10 & $< 1.1\times 10^{-6}$ & $<1.1 \times 10^{-4}$  \\
\mbox{\hspace{1.3in}} $\tan\!\beta$=50 & $< 1.3\times 10^{-6}$ & $<4.5 \times 10^{-6}$ \\
\mbox{\hspace{0.2in}} $\mu_{\scriptscriptstyle 0}=250$ GeV,
                      $\tan\!\beta$=2  & $< 1.7\times 10^{-6}$ & $<1.2 \times 10^{-2}$  \\
\mbox{\hspace{1.3in}} $\tan\!\beta$=10 & $< 2.4\times 10^{-6}$ & $<4.7 \times 10^{-4}$  \\
\mbox{\hspace{1.3in}} $\tan\!\beta$=50 & $< 2.7\times 10^{-6}$ & $<1.9 \times 10^{-5}$ \\
\mbox{\hspace{0.2in}} $\mu_{\scriptscriptstyle 0}=500$ GeV,
                      $\tan\!\beta$=2  & $< 3.9\times 10^{-6}$ & $<8.3 \times 10^{-2}$  \\
\mbox{\hspace{1.3in}} $\tan\!\beta$=10 & $< 6.4\times 10^{-6}$ & $<3.2 \times 10^{-3}$  \\
\mbox{\hspace{1.3in}} $\tan\!\beta$=50 & $< 7.5\times 10^{-6}$ &
$<1.3 \times 10^{-4}$
\end{tabular}
\end{table}
%%%%%%%%%%%%%%%%%%%%%%%%%%%%%%%%%%%%%%%%%%
\end{document}